\documentclass[useAMS,usenatbib]{mn2e}

\usepackage{natbib}
\usepackage{graphicx}
\usepackage{listings}
\usepackage[intlimits]{amsmath}
\usepackage{txfonts}
\usepackage{bm}
\usepackage{amssymb}
\usepackage{amsmath}
\usepackage{tabularx} 
\usepackage{color}

\newcommand{\vv}[1]{{\bmath #1}}
\newcommand{\oder}[2]{\frac{d #1}{d #2}}

\newcommand{\Pd}[1]{\partial_{#1}}
\newcommand{\Od}[1]{d_{#1}}

\newcommand{\ort}[1]{ \bmath{i}_{#1} }

\newcommand{\beq}{\begin{equation}}
\newcommand{\eeq}{\end{equation}}
\newcommand{\fracp}[2]{\left(\frac{#1}{#2}\right)}
\newcommand{\mtext}[1]{ \quad\mbox{#1}\quad}

\def\be{\begin{equation}}
\def\ee{\end{equation}}
\def\ba{\begin{eqnarray}}
\def\ea{\end{eqnarray}}
\def\go{\mathrel{\raise.3ex\hbox{$>$}\mkern-14mu
             \lower0.6ex\hbox{$\sim$}}}
\def\lo{\mathrel{\raise.3ex\hbox{$<$}\mkern-14mu
             \lower0.6ex\hbox{$\sim$}}}

\title[Magnetic  Centrifugal Instability]{Magnetic Inhibition of Centrifugal Instability}

\author[S.S. Komissarov, K.N. Gourgouliatos \& J. Matsumoto]{{Serguei S. Komissarov$^{1}$ \thanks{Email: S.S.Komissarov@leeds.ac.uk}, Konstantinos N. Gourgouliatos\thanks{Email: Konstantinos.Gourgouliatos@durham.ac.uk}$^{2}$ \& Jin Matsumoto$^{1,3}$ \thanks{Email: jin.matsumoto@fukuoka-u.ac.jp}}\vspace{0.4cm}\\
\parbox{\textwidth}{$^{1}$Department of Applied Mathematics, University of Leeds, Leeds LS2 9JT , UK }\\
\parbox{\textwidth}{$^{2}$Department of Mathematical Sciences, Durham University, Durham DH1 3LE, UK } \\
\parbox{\textwidth}{$^{3}$Research Institute of Stellar Explosive Phenomena, Fukuoka University, Fukuoka 814-0180, Japan }} 

\begin{document}

\date{Accepted -. Received -; in original form -}
\pagerange{\pageref{firstpage}--\pageref{lastpage}} \pubyear{-}
\maketitle

\label{firstpage}
    
\begin{abstract}
Recently it was shown that the centrifugal instability may be important in the dynamics of astrophysical jets undergoing reconfinement by external pressure. However, these studies were limited to the case of unmagnetised flows. Here we explore the role of the magnetic field within both the Newtonian and relativistic frameworks. Since the jet problem is rather complicated, we focus instead on the simpler problem of cylindrical rotation and axial magnetic, which shares significant similarity with the  jet problem, and consider only axisymmetric perturbations.   The studied equilibrium configurations involve a cylindrical  interface and they are stable to non-magnetic centrifugal instabilities everywhere except this interface. We use a heuristic approach to derive the local stability criterion for the interface in the magnetic case and numerical simulations to verify the role of the magnetic field. The theory and simulations agree quite well for Newtonian models but indicate a potential discrepancy for the relativistic models in the limit of high Lorentz factor of the rotational motion at the interface. In general, the magnetic field sets a critical wavelength below which the centrifugal modes are stabilised. We discuss the implication of our findings for the astrophysical jets, which suggest the centrifugal instability develops only in jets with relatively low magnetisation. Namely, the magnetic pressure has to be below the thermal one and for the relativistic case the jets have to be kinetic-energy dominated.           
 
\end{abstract}


\section{Introduction}
\label{Sec:Intro}

\begin{figure}
\includegraphics[width=1.0\columnwidth]{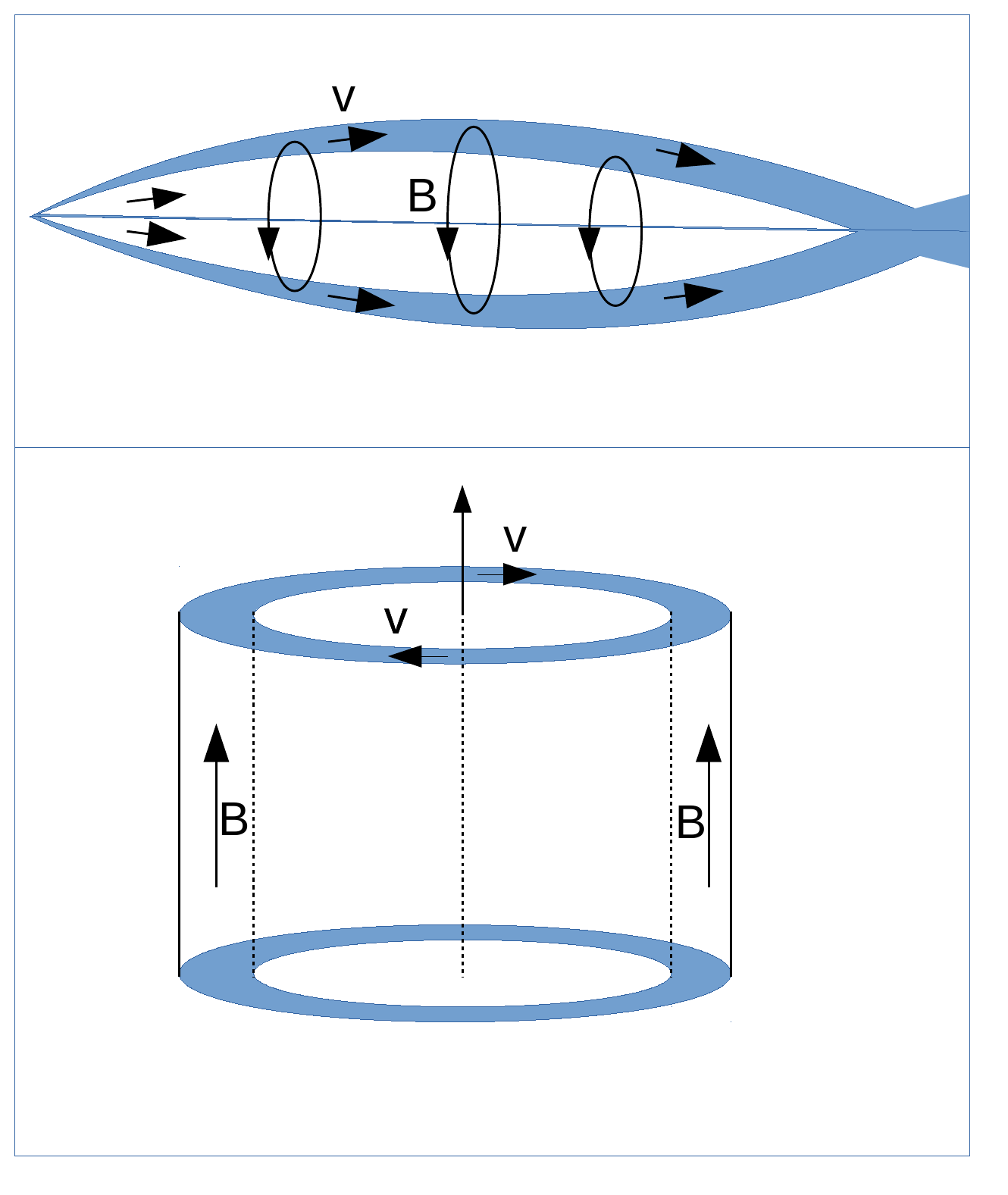}
\caption{{\it Top panel:} Jet with purely azimuthal magnetic field undergoing recollimation by external pressure. The white interior corresponds to the freely expanding unshocked flow. The shaded region corresponds to the shocked outer layer of the jet. The boundary between the two is the conical reconfinement shock driven into the jet by the external pressure. {\it Bottom panel:} The configuration describing a magnetised rotating cylindrical shell with purely axial magnetic field which is explored in this paper. }
\label{Fig:intro1}
\end{figure}

It is reasonable to assume that after initial collimation inside the central engine (black hole, its magnetised accretion disk and presumably a disk wind), AGN jets enter the phase of free expansion because of the fast decrease of the external pressure in the vicinity of AGN \citep[e.g.][]{PK-15}. During this phase the jets are globally stable and can extend well beyond AGN. However, at even larger distances the external pressure distribution is expected to flatten. This is the case inside the radio lobes (large-scale cocoons created by jets) and inside the cores of the hot gas found in their host galaxies. In these regions the external pressure may become important for the jet collimation once more, driving inside the jet a recollimation or a reconfinement shock.  While inside the jet the stream lines are straight, inside its  shocked outer layer they bend towards the jet axis (see Figure \ref{Fig:intro1}).  The 3D numerical simulations of unmagnetised jets undergoing the process of reconfinement have shown that such jets are susceptible to the centrifugal instability which facilitates a rapid transition to turbulence \citep{2018NatAs...2..167G}.  The unstable modes are non-axisymmetric and this is why the instability could not be observed in the previous axisymmetric simulations of  recollimated jets.                     

As the steady-state structure of reconfined jets is rather complicated and cannot even be found analytically, their stability cannot studied using the standard linear stability analysis.  Moreover, their 3D simulations are still computationally  expensive and do not allow a comprehensive numerical investigation of their instability. On the other hand, the centrifugal instability is local and for sufficiently small wavelengths does not depend on the details of the large-scale flow. For this reason \citet{2018MNRAS.475L.125G}  have turned their attention to rotating flows with axial symmetry.  They used heuristic arguments to generalise the Rayleigh instability criterion for incompressible flows to both compressible Newtonian and relativistic flows, both in the case of a continuous  profile of the rotational velocity and in the presence of tangential discontinuity. They also used computer simulations verifying the generalised Rayleigh criterion both for the Newtonian and relativistic cases.   

Many types of astrophysical jets are known to contain magnetic field as confirmed by Faraday Rotation measurements \citep{Asada:2002}.  Moreover, these jets are believed to be accelerated and collimated via the forces associated with the magnetic field anchored to the rotating central object. In the case of AGN jets, this is a supermassive black hole with its accretion disk  \citep{Blandford:1977, Blandford:1982}. Their rotation twists the magnetic field lines which develop characteristic helical structure.  In a steady-state axisymmetric jet, the longitudinal component of its magnetic field behaves as $B_\parallel \propto r_j^{-2}$ and the azimuthal one as $B_\phi\propto r_j^{-1}$,  where $r_j$ is the jet radius. For this reason, the azimuthal field is expected to dominate at the large distances from the central engine where the jets are expected to undergo reconfinement and recollimation.  For such a magnetic configuration the CFI-unstable modes would corrugate the magnetic field lines, thus increasing the magnetic energy.   Hence the magnetic field should have a stabilising effect, like it does in the magnetic Rayleigh-Taylor instability \citep{Chandra1961}.  In the context  of the stability of viscous flow between two rotating cylinders the inhibiting effect  of magnetic field was studied theoretically by \citet{Chandra-53} and experimentally by \citet{DO-60,DO-62}.  Recently, \citet{BMRM-19} carried out the linear stability analysis of rotating cylindrical relativistic jets with vanishing gas pressure. These jets are unbounded and their equilibrium requires dynamically important poloidal component of the magnetic field. They identify CFI modes driven by the rotation and describe their stabilisation by the magnetic field.    

In this paper, we follow  \citet{2018MNRAS.475L.125G} and use a rotating cylindrical configuration in order to study the role of the magnetic field. The problem setup is outlined in the bottom panel of Figure~\ref{Fig:intro1}.  The magnetic field is aligned with the axis of rotation as in this case it is normal to the streamlines and parallel to the interface with the external gas at rest, just like in the jet configuration shown in the top panel of Figure~\ref{Fig:intro1}. The flow is stable to CFI both above and below the interface and only the interface can develop the instability.  This simplifies the analysis and allows a straightforward interpretation of the results. 

This configuration is also relevant for the magnetorotational instabilitly (MRI) \citep{Velikhov:1959,Chandrasekhar:1960, Balbus:1991}, that has been recognised as very important in astrophysics in the context of the accretion disc dynamics.  We note that the difference between our setup and the MRI is that here we consider a fluid flow that is unstable under the centrifugal instability and we explore the conditions under which it can be stabilised under the presence of a magnetic field. The MRI setup, on the contrary, is stable under the centrifugal instability and is destabilised through the presence of a magnetic field. 

The plan of the paper is as follows. In section \ref{Sec:Heuristic} we present the heuristic derivation of the instability criterion for such a configuration, both in the Newtonian and relativistic frameworks.   In section \ref{Sec:simulations-overview} we provide an overview of our computer simulations, designed to study the actual properties of the magnetic CFI and to test its heuristic criterion.  The detailed initial setup and results of the Newtonian simulations are described in section~\ref{Sec:cs-nc}, whereas section \ref{Sec:cs-rc} describes the relativistic simulations.  In section \ref{Discussion} we discuss the results of our study and their implications to astrophysical flows and  our  conclusions are summarised  in section \ref{Conclusions}.
%

\section{Heuristic instability criterion} 
\label{Sec:Heuristic}

The steady state configuration considered in our study involves an axisymmetric rotating fluid confined by two cylindrical 
walls, the inner one at the radius $r_1=r_{in}-\Delta r$ and the outer one at the radius $r_2=r_{in}+\Delta r$, where $r_{in}$ is radius of the interface, which is unstable to CFI in the case of vanishing the magnetic field. In the cylindrical coordinates $\{r,\phi,z\}$ aligned with symmetry axis of the problem, the velocity $\vv{v}= v^\phi(r) \ort{\phi}$ and the magnetic field $\vv{B}=B^z(r)\ort{z}$. The fluid density $\rho=\rho(r)$ and pressure $p=p(r)$.   For simplicity, we further assume that not only the total pressure, but also its thermal and hence magnetic pressure is continuous across the interface.      

\subsection{Newtonian case} 
\label{Subsec:HN}

Using the Einstein summation convention, the continuity equation of Newtonian fluid dynamics and MHD can be written as
\be 
\Pd{t} \rho +\nabla_i (\rho v^i) = 0 \,, 
\label{hn-con}
\ee
where  $v^i$ is the contravariant component of the velocity vector and $\nabla_i$ is the covariant derivative operator of  Euclidean space. Similarly, the momentum equation can be written as 
\be 
\Pd{t} \left(\rho v_j\right) +\nabla_i (\rho v^i v_j + p\delta^i_j) +\nabla_i \left( \frac{B^2}{8\pi} \delta^i_j 
- \frac{B^i B_j} {4\pi}\right) = 0  \,,
\label{hn-mom}
\ee
where $B^i$ and $B_i$ are the contravariant and covariant components of the magnetic field vector and $v_i$ are the covariant components of the velocity. The magnetic term of this equation can be expanded as 
\be 
\nabla_i \left( \frac{B^2}{8\pi} \delta^i_j - \frac{B^i B_j} {4\pi}\right) = 
(\delta^k_j - \hat{s}_j \hat{s}^k) \Pd{k} \fracp{B^2}{8\pi}  - \frac{B^2}{4\pi R_c} \hat{n}_j  \,,
\label{mag-term-n}
\ee
where $\hat{\vv{s}}$ is the unit vector tangent to the magnetic field line ($B^i = B \hat{s}^i$), $R_c$ is the curvature radius of the line and $\hat{\vv{n}}$ is the unit vector normal to  $\hat{\vv{s}}$ and pointing towards its centre of curvature. Hence 
the momentum equation reads 
\be 
\Pd{t} \left(\rho v_j \right)+\nabla_i (\rho v^i v_j)  + \nabla_j(p+p_m)  - \hat{s}_j \hat{s}^k \nabla_k p_m - 
\frac{B^2}{4\pi R_c} \hat{n}_j  = 0  \,.
\label{hn-mom1}
\ee
Replacing the covariant divergences of vector and symmetric tensor fields in Eq.~\ref{hn-mom1} and Eq.~\ref{hn-con} with their expressions in terms of the coordinate derivatives \citep{Landau:1971}  and combining the two one obtains the
equation of motion   
\be 
\rho\,\Od{t} v_j = - \Pd{j} p_{tot}  + \rho\frac{v^k v^l}{2} \Pd{j} g_{kl} +  \frac{B^2}{4\pi R_c} \hat{n}_j +  \hat{s}_j \hat{s}^k \Pd{k} p_m  \,,
\label{hn-mom2}
\ee
where $\Pd{i}\equiv \partial/\partial x^i$,  $\Od{t} = \Pd{t} + v^i\Pd{i}$, $p_{tot}=p+p_m$ and $g_{kl}$ are the covariant components of the metric tensor of Euclidean space.   

In the cylindrical coordinates, the radial component of Eq.~\ref{hn-mom2} reads 
\be 
\rho\,\Od{t}v_{\hat{r}} = - \Pd{r} p_{tot} + \rho\frac{v_{\hat{\phi}}^2}{r} +  \frac{B^2}{4\pi R_c} \hat{n}_r +  \hat{s}_r \hat{s}^k \Pd{k} p_m  \,,
\label{hn-mom-r1}
\ee
where $v_{\hat{i}}$ are the velocity components in the normalised coordinate basis.  Hence in the steady state we have 
\be 
- \Pd{r} p_{tot}  + \rho\frac{v_{\hat{\phi}}^2}{r}  = 0  \,. 
\label{equilib-n}
\ee

Here we analyse the stability of a cylindrical interface where $\rho (v^{\hat{\phi}})^2$ experiences finite jump. The flow parameters below the interface we will denote using index "1" and those above it using index "2".    
Consider an axisymmetric perturbation of the interface in the form of a small ring that slowly rises across the interface. After the crossing, its pressure distribution adjusts to that of the surrounding fluid and it becomes subject to the finite radial force per unit volume 
\beq
     f_{d} =  -\frac{1}{r_{in}}[\rho v_{\hat{\phi}}^2]  \,,
     \label{driving}
\eeq
where $[A]=A_2-A_1$ is the jump across the interface  \citep{2018MNRAS.475L.125G}.  
For unmagnetised fluids this leads to the instability criterion  
\beq
     [\rho v_{\hat{\phi}}^2] <0 \,.
\eeq

Out of the two terms in Eq.~\ref{hn-mom-r1} attributed to the magnetic tension only  
\beq
     f_m = \frac{B^2}{4\pi R_c} \hat{n}_r   
     \label{m-tension}
\eeq
is clearly a restoring force. The term $\hat{s}_r \hat{s}^k \Pd{k} p_m$ simply balances the magnetic pressure gradient along the magnetic field line. Moreover, in the linear stability analysis it would yield only a second order term as in the steady state $\hat{s}_r=0$ and only $\Pd{r}\not=0$.  For these reasons we will ignore this component.   

If the magnetic tension (\ref{m-tension}) becomes comparable to the driving force (\ref{driving}) already for $R_c$ much higher than the length scale of the perturbation (small curvature of the magnetic field lines) then it seems natural to expect that the instability will be suppressed and only a small amplitude oscillation will be produced instead. $R_c$ smaller or comparable to the length scale of the perturbation means strong deformation of the magnetic field lines and if the balance between the magnetic tension and the driving force requires such a deformation then the instability is unlikely to be suppressed.  This physical argument implies the existence of critical wavelength
$$
   \lambda_c= \alpha R_{c}^* \,,
$$   
where $R_{c}^*$ is the curvature radius corresponding to the balance between the driving and restoring forces, 
\beq
       \frac{1}{R_{c}^*}\frac{B^2}{4\pi} =  -\frac{1}{r_{in}}[\rho v_{\hat{\phi}}^2] \,.
       \label{ring-balance-n}
\eeq
Obviously, the instability is suppressed for modes with $\lambda<\lambda_c$ and is allowed to develop when  $\lambda>\lambda_c$

Provided $\alpha$ is constant,  its value can be found using the identity between CFI with continuous velocity across the interface and the Rayleigh-Taylor instability \citep{2018MNRAS.475L.125G}.  In the case of incompressible fluid,  the critical wavelength of the magnetic Rayleigh-Taylor instability is 
\beq
      \lambda_c = \frac{B^2}{g(\rho_1-\rho_2)} 
      \label{eq-b}
\eeq 
(Chandrasekhar 1961) whereas for $[v_{\hat{\phi}}]=0$  equation (\ref{ring-balance-n}) yeilds 
\beq
      \lambda_c = \alpha \frac{B^2}{4\pi} \frac{1}{g(\rho_1-\rho_2)} \,,
      \label{eq-c}
\eeq 
where $g=v_{\hat{\phi}}^2/r_{in}$ is the centrifugal acceleration. These two expressions match when 
$\alpha=4\pi$, suggesting that the critical wavelength of the magnetic centrifugal instability is 
\beq  
       \lambda_c = \frac{B^2}{ \rho_1 v_{\hat{\phi},1}^2 - \rho_2 v_{\hat{\phi},2}^2} r_{in} \,. 
      \label{lambda-cr-n}
\eeq  

Shivamoggi (1982) has shown that compressibility introduces corrections of the order of 
$\beta^{-1} = p_m/ p$ to the equations of the incompressible magnetic RTI. Already for this reason we expect Eq.~\ref{lambda-cr-n} to be only approximate. Yet, in the absence of proper linear stability analysis,  it provides a reasonable reference point against which the results of our numerical study can be compared.

\subsection{Relativistic case} 
\label{Subsec:HR}

In the 3+1 form, the relativistic continuity equation is 
\beq
 \Pd{t} \left(\rho u^t \right) +\nabla_i \left(\rho u^i \right)= 0  \,,
\eeq 
where $\rho$ is the rest mass-energy of the fluid and $u^\nu$ is its 4-velocity vector. 
The energy-momentum equation is 
\beq
\Pd{t} T^t_\mu+\nabla_i T^i_\mu = 0 \,. 
\eeq
where 
\beq
 T^\nu_\mu = (w+2p_m)u^\nu u_\mu + (p+p_m) \delta^\nu_\mu - \frac{b^\nu b_\mu}{4\pi}  
\eeq
is the stress-energy-momentum tensor of magnetised fluid,  $w=\rho c^2 + \gamma p/ (\gamma-1)$ is the relativistic enthalpy, $p$ is the thermodynamic pressure, $b^\nu$ is the 4-vector of magnetic field and  $p_m=b^2/8\pi$ is the magnetic pressure \citep{Dixon:1978,Anile:1989}. The standard 3-vector of magnetic field is related 
to the components of $b^\nu$ via 
\beq
   b^t=(u_i B^i) \mtext{and}   b^j=\frac{1}{u^t} B^j + b^t v^j \,.
\eeq 
Hence the momentum equation is  
\beq
\Pd{t} (\tilde{w} u^t u_j) + \nabla_i (\tilde{w} u^i u_j) + \nabla_j (p_{tot}) - 
\frac{1}{4\pi} (b_j \nabla_\nu b^\nu  + b^\nu \nabla_\nu b_j )=0\,, 
\label{hr-mom}
\eeq
where $\tilde{w} =w + 2p_m$.  

Proceeding by replacing the covariant derivatives with the coordinate derivatives and using the continuity equation in the same fashion as in the Newtonian derivation, one finds the radial component of the equation of motion 
\beq 
\rho\, \Od{\tau} (\tilde{h} u_r) = \frac{\tilde{w} u_\phi^2}{2} - \Pd{r} p_{tot} +
\frac{1}{4\pi} (b_r \nabla_\nu b^\nu  + b^\nu \nabla_\nu b_r )\,, 
\eeq
where $\tilde{h}=\tilde{w}/\rho$,  $\Od{\tau}=u^\nu \Pd{\nu}$ and hence $\tau$ is the proper time of the fluid element.  
This is the relativistic counterpart of equation~(\ref{hn-mom-r1}). In the steady-state configuration $b^r=0$ and hence  
\beq 
\frac{\tilde{w} u_\phi^2}{2} - \Pd{r} p_{tot} =0 \,.
\label{equilib-r}
\eeq
This equation generalises equation~(\ref{equilib-n}) of the Newtonian case.

Here we argue that for the problem in hand one may put  
\beq
b_r \nabla_\nu b^\nu  + b^\nu \nabla_\nu b_r \simeq \frac{b^2}{R_c}  \hat{n}_r \,,
\eeq
where $R_c$ is the local curvature radius of the field line of the 3-vector field $\vv{b}$, defined via $b^\nu=(b^t,\vv{b})$.  Indeed, 
$$
  b_r \nabla_\nu b^\nu = b_r \Pd{t} b^t + b_r \nabla_i b^i 
$$ 
and since in the steady state $b_r=0$ and $\nabla_i b^i=0$  this term is at least second order.  We also have 
$$
  b^\nu \nabla_\nu b_r = b^t \Pd{t} b_r + b^i\nabla_i b_r = 
  b^t \Pd{t} b_r + \frac{\tilde{b}^2}{R_c} \hat{n}_r +  \hat{s}_r \hat{s}^k \Pd{k} \frac{\tilde{b}^2}{2} \,,
$$
where $\tilde{b}^2=b_ib^i = b^2-b_tb^t$. The first term on the right is at least second order because $b^t=0$ in the steady state. The last term is also at least second order because in the steady state $s^r=0$ and only $\Pd{r}\not=0$.  
In the remaining term one may replace $\tilde{b}^2$ with $b^2$ because $b_t=0$ in the steady state. 

Proceeding along the same line of arguing as in newtonian case we then conclude that the outcome is set by the competition between the instability-driving force 
\beq
     f_d = -\frac{1}{r_{in}}[\tilde{w} u_{\hat{\phi}}^2]
\eeq
and the magnetic tension force 
\beq
  f_m = \frac{1}{R_{c}}\frac{b^2}{4\pi}  \,.
\eeq
This defines the critical wavelength of the instability  
\beq  
       \lambda_c = \alpha \frac{b^2}{ 4\pi (\tilde{w}_1 u_{\hat{\phi},1}^2 - \tilde{w}_2 u_{\hat{\phi},2}^2) } r_{in} \,. 
      \label{lambda-cr-r}
\eeq  
In the Newtonian limit, this reduces to the result (\ref{lambda-cr-n}) only if $\alpha=4\pi$. 

\section{Computer simulations. Overview}
\label{Sec:simulations-overview}

To probe the actual impact of magnetic field on the centrifugal instability we have carried out a rather comprehensive numerical study via axisymmetric MHD simulations. Both the Newtonian and relativistic cases were investigated. In the simulations, we used the equation of state of ideal gas with the ratio of specific heats $\gamma=5/3$ for the Newtonian runs and $\gamma=4/3$ for the relativistic runs.

The simulations were carried out with AMRVAC code \citep{2012JCoPh.231..718K}, using HLLC Riemann solver \citep{Harten:1997},  Koren flux limiter \citep{Koren:1993} and three-step method for time integration. GLM approach was used to keep the magnetic field divergence free \citep{Dender:2002}.

\begin{figure}
\includegraphics[width=0.9\columnwidth]{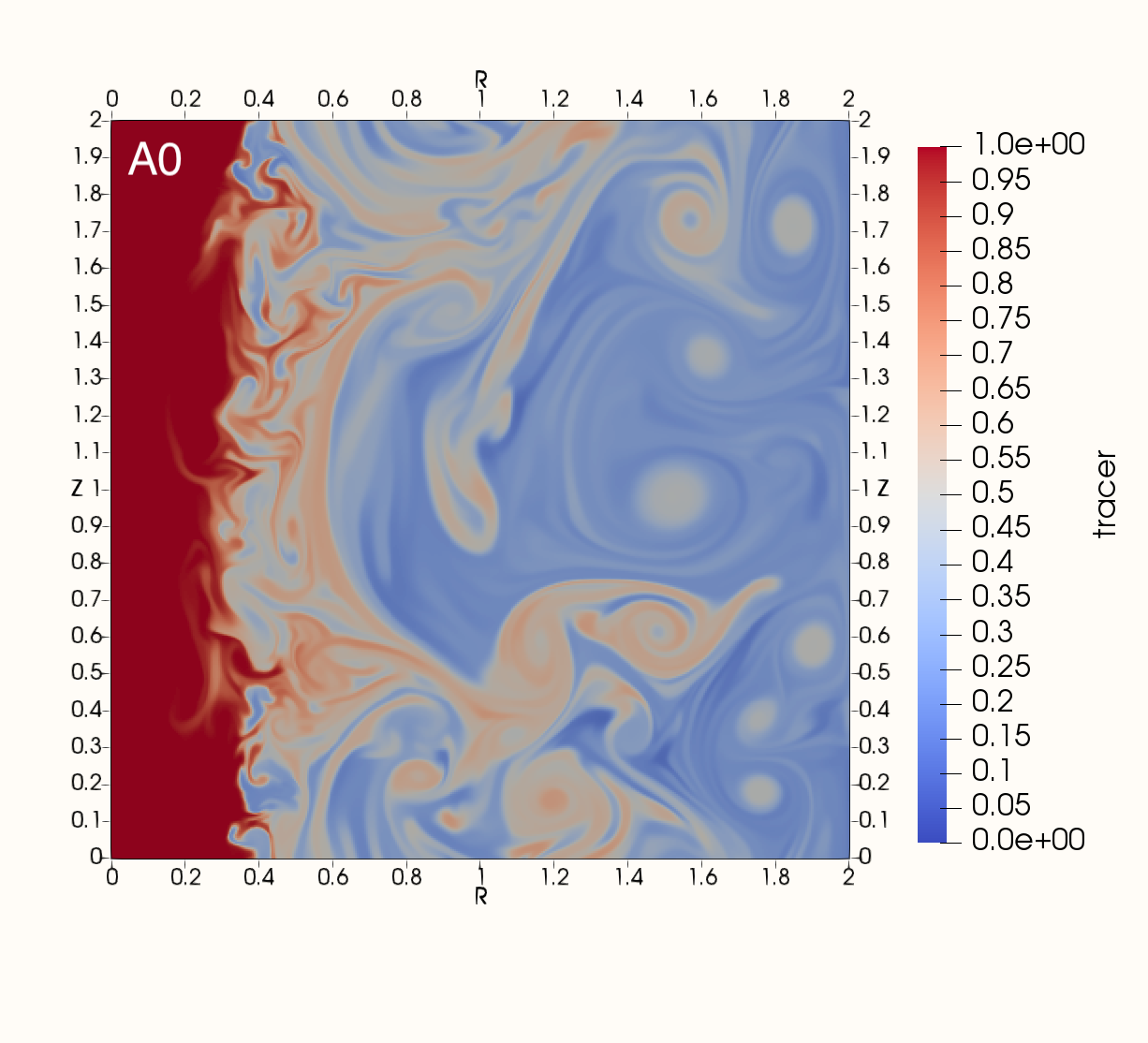}
\includegraphics[width=0.9\columnwidth]{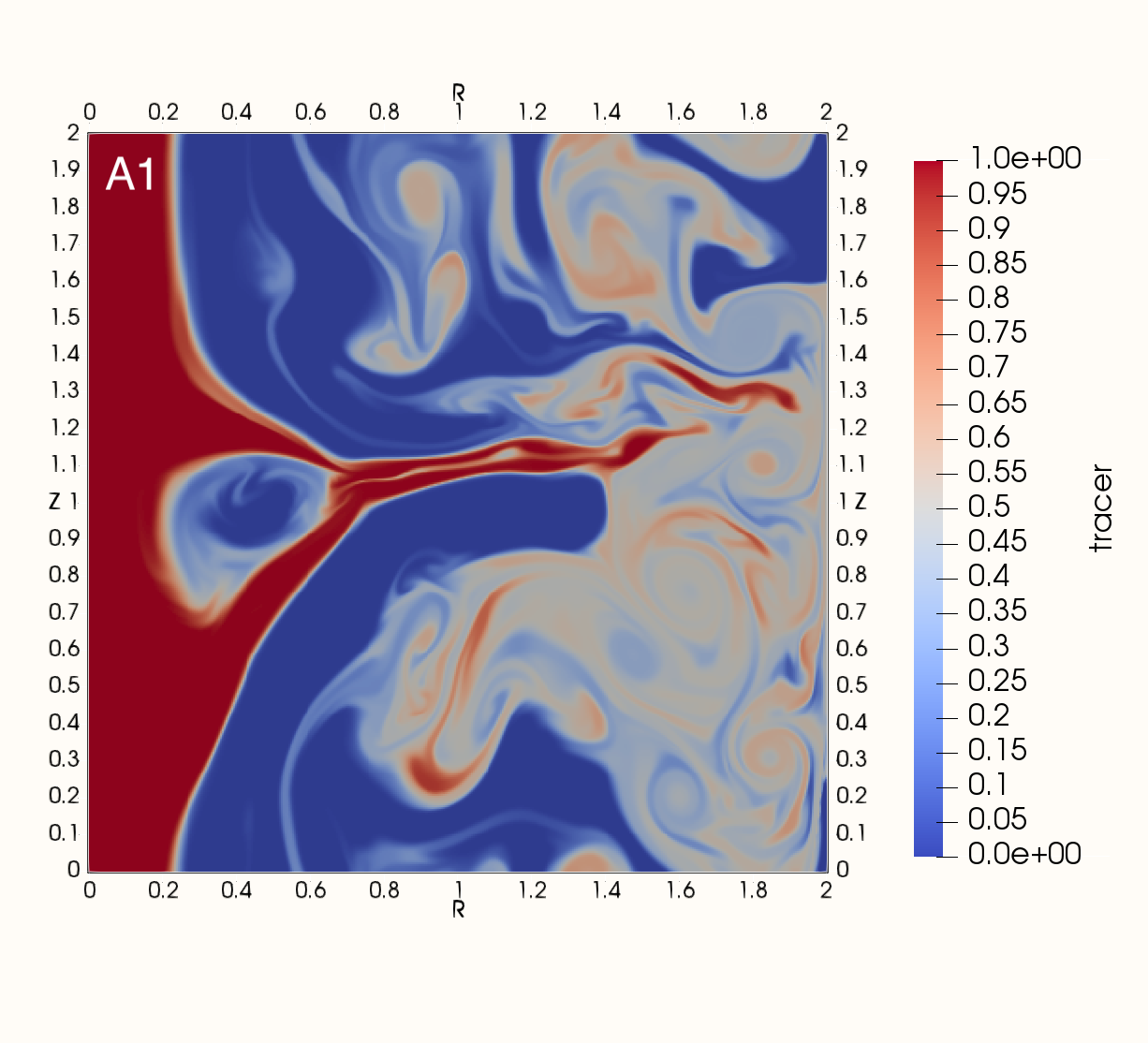}
\includegraphics[width=0.9\columnwidth]{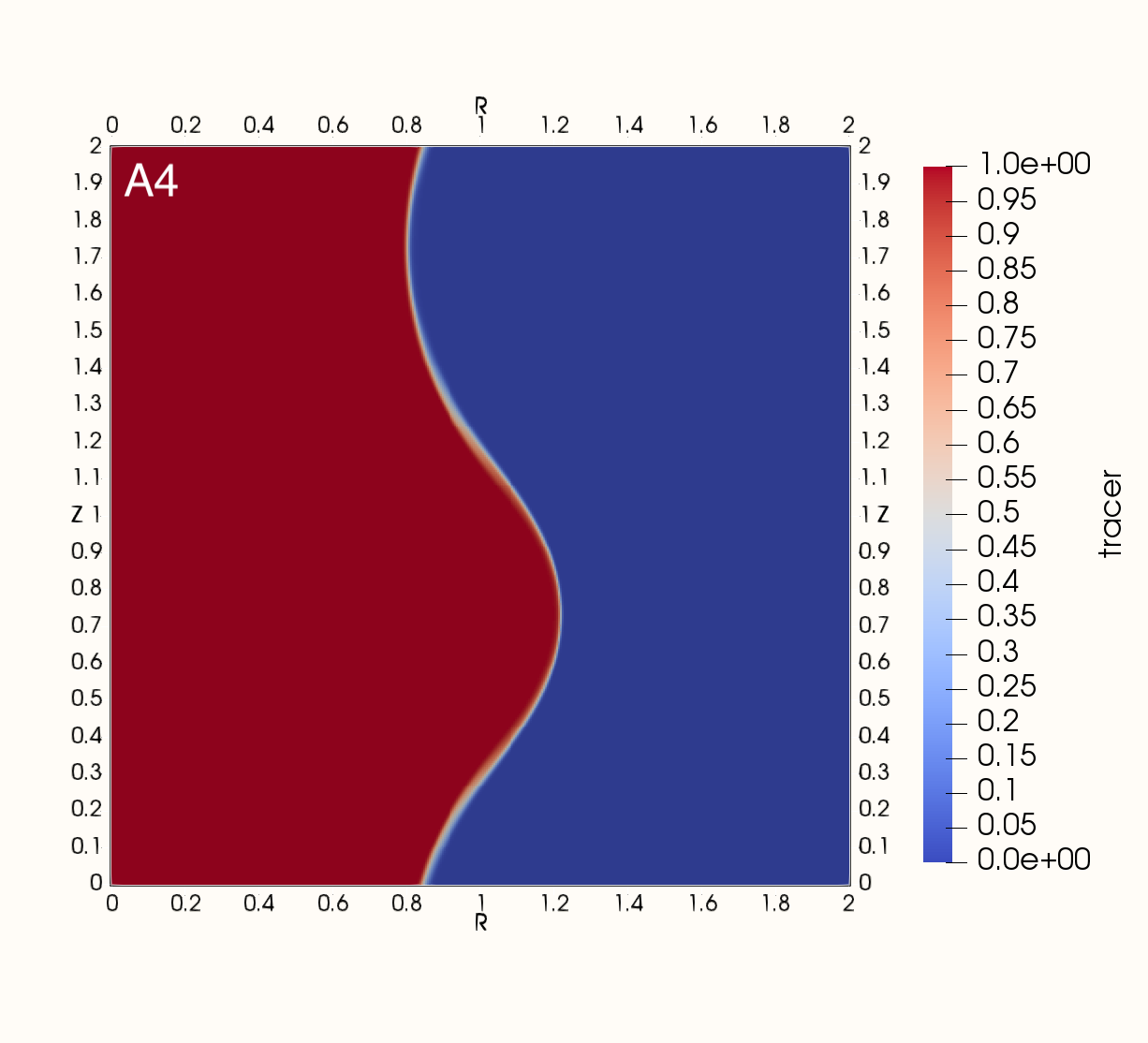}
\caption{Snapshots of the Newtonian solutions in the case of rotating cylinder.  The three panels represent the models A0 (top), A1 (middle) and A4 (bottom) at the time $t=20$, where the instability has grown to a non-linear level in all three cases. The shown parameter is the advective tracer.}
\label{Fig:Newton1}
\end{figure}

The main initial configuration involves a thin rotating fluid shell surrounded by a fluid at rest on the outside and a solid wall on the inside. Hence the computational domain is $r\times z \in \left[r_{in}-\Delta r, r_{in}+\Delta r \right]\times\left[0, \Delta z\right]$, where $r_{in}$ is the radius of the interface between the fluids and $\Delta r \ll r$ is the shell thickness.  We set the dimensionless  $r_{in}=1$. 

The values of $\Delta r$ and $\Delta z$ can be selected based on the analogy of the rotating shell and reconfined jet as discussed in the Introduction. Approximating the jet boundary by an arc of radius $r_{in}$ and using the small angle approximation, we find        
\beq
     r_j \simeq \frac{1}{2}\theta_j^2 r_{in} \, ,
     \label{rjc-radius}
\eeq
where $\theta_j$ is the jet's initial half-opening angle and $r_j$ is its radius at the bulge. 
For $r_{in}=1$ and the reasonable value $\theta_j=0.1$ we have $r_j \simeq 0.005$. Since the width of the shocked outer layer of a reconfined jet cannot exceed it bulge radius, it makes sense to put $\Delta r \lesssim 0.005$ as well.   
 
The $z$ direction in the problem of cylindrical rotation corresponds to the azimuthal direction in the problem of reconfined jet. Hence the wavelength $\lambda$ in the rotational case roughly corresponds to $2\pi r_j/m$ in the jet problem, where $m=1,2,\dots$ is the azimuthal wave number, and so with application to the jet problem only $\lambda< 2\pi r_j$ are of importance.  Hence we may put   
\begin{eqnarray}
       \Delta z \lesssim \pi \theta_j^2  r_{in} \,. 
\end{eqnarray}
 For $r_{in}=1$ and $\theta_j=0.1$ we obtain $\Delta z \lesssim 0.03$. These values for $\Delta r$ and $\Delta z$ are reasonable estimates rather than strict constraints and this is how we utilise them in the study. Below we describe an alternative way of setting $\Delta r$, which gives similar values and was actually utilised in the simulations.

Because of the local nature of CFI, one expects the shell configuration to yield generic conclusions. To make sure that this indeed the case and provide continuity with our study of the unmagnetised case \citep{2018MNRAS.475L.125G}, we also considered the setup where the shell is replaced with a whole cylinder rotating uniformly ($\Delta r=r_{in}$). This was done only in the Newtonian case. The exact solutions used to setup initial configurations are described later in the sections dedicated to the individual cases.

In the $z$ direction we used periodic boundary conditions for all variables. For the shell configuration,  we imposed antisymmetric boundary conditions for the radial components the velocity and magnetic field and symmetric ones for all other quantities at both the radial boundaries. For the cylinder configuration we used the same boundary conditions as in the shell configuration for all variables, with the exception of the azimuthal components of the velocity and magnetic field, for which we the imposed the antisymmetry condition instead.

As a rule, we used a uniform computational grid with square cells, with 400 of them in the radial direction.
For some models, we used higher resolution to check the convergence and to make sure that location of the transition from stable to unstable regimes is not influenced by the grid. In few cases, where a particularly high resolution was needed to resolve the thin boundary layer of the initial configuration and hence reduce the numerical dissipation, we used the adaptive mesh facility of the code (AMR), with up to three levels of refinement. 

\begin{table*}
	\centering
	\caption{ Parameters of Newtonian runs for the case of rotating cylinder. $\rho$ - the mass density,  $\Omega$ - the angular velocity, $p_0$ - the pressure at the symmetry axis, $B$ - the strength of the axial magnetic field, $M_{a}$ - the Alfv\'enic Mach number at the interface, $M_s$ - the usual Mach number at the interface, $\beta$ - the magnetisation at the interface,  S/U- the stability indicator, $r\times z$ - the integration domain, $t_f$ - the total time of the run.}
	\label{tab:C}
	\begin{tabular}{lcccccccccccccccc} 
		\hline
		name &$\rho_1$ & $\rho_2$ & $\Omega_1$ & $\Omega_2$ & $p_0$ & $B$&$M_{a,1}$&$M_{a,2}$& $M_{s,1}$ & $M_{s,2}$ & $\beta_{in}$& S/U  &grid & $r\times z$ & $t_f$\\
		\hline
		A0 & 2 & 1 & 1 & 0& 11&    0& 0  &0 &$0.31$& $0$&  $\infty$ & U & $400^2 $ & $[0,2]\times [0,2]$ & 20.\\
		A1 & 2 & 1 & 1 & 0& 11 &0.01& 141&0 &$0.31$& $0$&  $2.4\times10^5$ & U & $400^2 $& $[0,2]\times [0,2]$ & 20.\\
		A2 & 2 & 1 & 1 & 0& 11 &0.1& 14  &0 &$0.31$& $0$&  $2.4\times10^3$ & U & $400^2 $& $[0,2]\times [0,2]$ & 20.\\
		A3 & 2 & 1 & 1 & 0& 11 &0.2& 7   &0 &$0.31$& $0$&  $6\times10^2$ & U & $400^2 $& $[0,2]\times [0,2]$ & 20.\\
		A4 & 2 & 1 & 1 & 0& 11 &0.5& 2.8 &0 &$0.31$& $0$&  96& U & $400^2 $& $[0,2]\times [0,2]$ & 20.\\
		A5 & 2 & 1 & 1 & 0& 1.1 &0.5& 2.8&0 &$0.76$& $0$&  16.8 &  U & $400^2 $& $[0,2]\times [0,2]$ & 20.\\
		A6 & 2 & 1 & 1 & 0& 11 &1.0& 1.4 &0  &$0.31$& $0$& 24&  S & $400^2 $& $[0,2]\times [0,2]$ & 20.\\
	 	A7& 2 & 1 & 1 & 0& 1.1 &1.0& 1.4 &0  &$0.76$& $0$& 4.2& S & $400^2 $& $[0,2]\times [0,2]$ & 20.\\
		\hline
		B1 & 2 & 1 & 1 & 1& 11.0 &0.1& 14&10 &$0.31$& $0.22$& $2.4\times10^3$ &   U & $400^2 $& $[0,2]\times [0,2]$ & 20.\\
		B2 & 2 & 1 & 1 & 1& 11.0 &0.2& 7 &5  &$0.31$& $0.22$& $6\times10^2$ &   U & $400^2 $& $[0,2]\times [0,2]$ & 20.\\
		B3& 2 & 1 & 1 & 1& 11.0 &0.5& 2.8& 2 &$0.31$& $0.22$& 96&   S & $400^2 $& $[0,2]\times [0,2]$ & 20.\\
		\hline
		C1 & 2 & 1 & 2 & 1& 14 &0.5& 5.7 &2  &$0.51$& $0.18$& $1.4\times10^2$&   U & $400^2 $& $[0,2]\times [0,2]$ & 20.\\
		C2 & 2 & 1 & 2 & 1& 14 &1.0& 2.8 &1  &$0.51$& $0.18$& 36&   U & $400^2 $& $[0,2]\times [0,2]$ & 20.\\
		C3& 2 & 1 & 2 & 1& 14 &1.5& 1.9 &0.7 &$0.51$& $0.18$& 16&   S & $400^2 $& $[0,2]\times [0,2]$ & 20.\\
      		\hline
	\end{tabular}
\end{table*}

In all simulations we introduce a small perturbation consisting of a several modes ($n_{max}>5$), where the longest wavelength mode has the size of the computational domain. The velocity perturbation is superposed to the initial azimuthal velocity $v_{\phi}(t=0)$:
\begin{eqnarray}
\delta v_{\phi}= a_{pert} \left(\frac{r-r_0}{\Delta r}\right)\sum_{n=1}^{n_{max}} (-1)^n \sin\left( \frac{2\pi n z}{\Delta z}\right)\,,
\end{eqnarray}
where  $r_0=r_{in} - \Delta r$ is the inner radius of the domain and  $\Delta z$ is its vertical size.   The amplitude of the perturbation is set to $a_{pert}<10^{-3}$ in all runs.  

An equilibrium state is considered unstable if the perturbations grow to non-linear levels. We set as a conventional criterion that the interface is displaced at least at a distance of $0.1\Delta r$ from its original position. In the simulations performed we have not found any borderline cases, i.e.~displacements marginally below or above this conventional limit. The simulation duration is always longer than one rotation period of the unperturbed flow at $R_1$. If the instability has grown to non-linear levels confirming that the configuration is unstable, the simulations is stopped. 

In order to help with the identification of the fluids initially located at the different sides of the interface, we introduced an advective tracer $\tau_p$, which is initially set to the value of 1 inside the interface and to the value of 0 outside of it. In the relativistic simulations its evolution is governed by the transport equation 
\begin{eqnarray}
\partial_{t}(\rho \tau_p u^t)+\nabla_{i}(\rho \tau_p u^i)=0 \,, 
\end{eqnarray}
whereas in the Newtonian simulations by 
\begin{eqnarray}
\partial_{t}(\rho \tau_p)+\nabla_{i}(\rho \tau_p v^i)=0 \,.
\end{eqnarray}

\section{Computer simulations. The Newtonian case}
\label{Sec:cs-nc}

\subsection{Rotating cylinder}
\label{Sec:cs-nc-rc}

In this configuration, the inner fluid occupies  
 $0<r<r_{in}$ and rotates with constant angular velocity $\Omega_1$. Its density distribution is uniform, $\rho=\rho_1$.  The outer fluid occupies $r_{in}<r<2r_{in}$ and rotates with angular velocity $\Omega_2$. Its density distribution is also uniform, $\rho=\rho_2$.  The magnetic field is parallel to the symmetry axis and uniform throughout the entire domain. The pressure distribution has to satisfy the equilibrium equation
\be 
- \Pd{r} p_{tot}  + \rho\frac{v_{\hat{\phi}}^2}{r}  = 0  
\label{equilib-n1}
\ee
(see Sec.~\ref{Subsec:HN}), which yields 
\begin{eqnarray}
    p = \left\{ 
    \begin{array}{cc}
         p_0+\frac{1}{2}\rho_1\Omega_1^2 r^2 & r \leq r_{in} \\
         p_{in}+\frac{1}{2}\rho_2\Omega_2^2(r^2-r_{in}^2) & r>r_{in}\,, 
    \end{array} \right.
\end{eqnarray}
where $p_0$ is the central pressure and $p_{in} = p_0+\frac{1}{2}\rho_1\Omega_1^2 r_{in}^2$ is the pressure at the interface between the fluids. 
The density and angular velocity profiles are chosen so that the quantity $\Psi= \rho \Omega^2 R^4$ reduces at the interface and so without the magnetic field that the system would suffer the centrifugal instability \citep{2018MNRAS.475L.125G}.  

In the non-relativistic physics, the degree of plasma magnetisation of magneto-static configurations is traditionally described with the dimensionless parameter $\beta=p/p_m$. It differs from the ratio of the thermal and magnetic energy densities by a factor of order unity.  For dynamic configuration, the Alv\'enic Mach $M_a=v/c_a$, where $c^2_a=B^2/4\pi\rho$, is another dimensionless parameter that can be useful to describe the magnetisation. Indeed, $M_a^2$ is twice the ratio of the kinetic energy of bulk motion and the magnetic energy.  We use both these parameters, as measured at the interface, to parametrise our numerical models.      
These and other parameters are shown in Table~\ref{tab:C}. 

The numerical models of this configuration split into three groups. In the group A, the external fluid is not rotating and in the unmagnetised limit this configuration is unstable to CFI irrespectively of the density jump at the interface. In the group B, both the fluids rotate with the same angular velocity. This configuration is analogous to the inertial RTI and hence we put $\rho_1>\rho_2$ to drive the instability. In the group C, the angular velocity of the inner fluid is increased compared to the group B, making the conditions even more favourable for CFI.    

Figure~\ref{Fig:Newton1} shows the final solutions for three models of the group A with progressively increasing magnetisation degree, namely A0, A1 and A4.  Their structure is consistent with the anticipation that the magnetic field suppresses  short-wavelength perturbations. Indeed in the model A0  the magnetic field vanishes ($\beta=\infty$ ) and solution exhibits much fine structure.  In the model A1, the magnetic field does not vanish and although it is rather weak ($\beta=2.2\times 10^4$ at the interface ), it erases most of the fine structure. In the model A4 with $\beta=88$ at the interface only the mode with the wavelength equal to the $z$ size of the domain can be seen.

All but two models in the group A have the same gas pressure at the interface and hence the Alfv\'enic Mach number $M_{a1}$ monotonically decreases with $\beta$. To break this degeneracy we introduce models A5 and A7, which have ten times lower gas pressure and hence $\beta$ at the interface.  The combined results suggest that whether stability develops  or not depends rather on  $M_{a1}$ than on $\beta$.

\begin{figure}
\includegraphics[width=1.0\columnwidth]{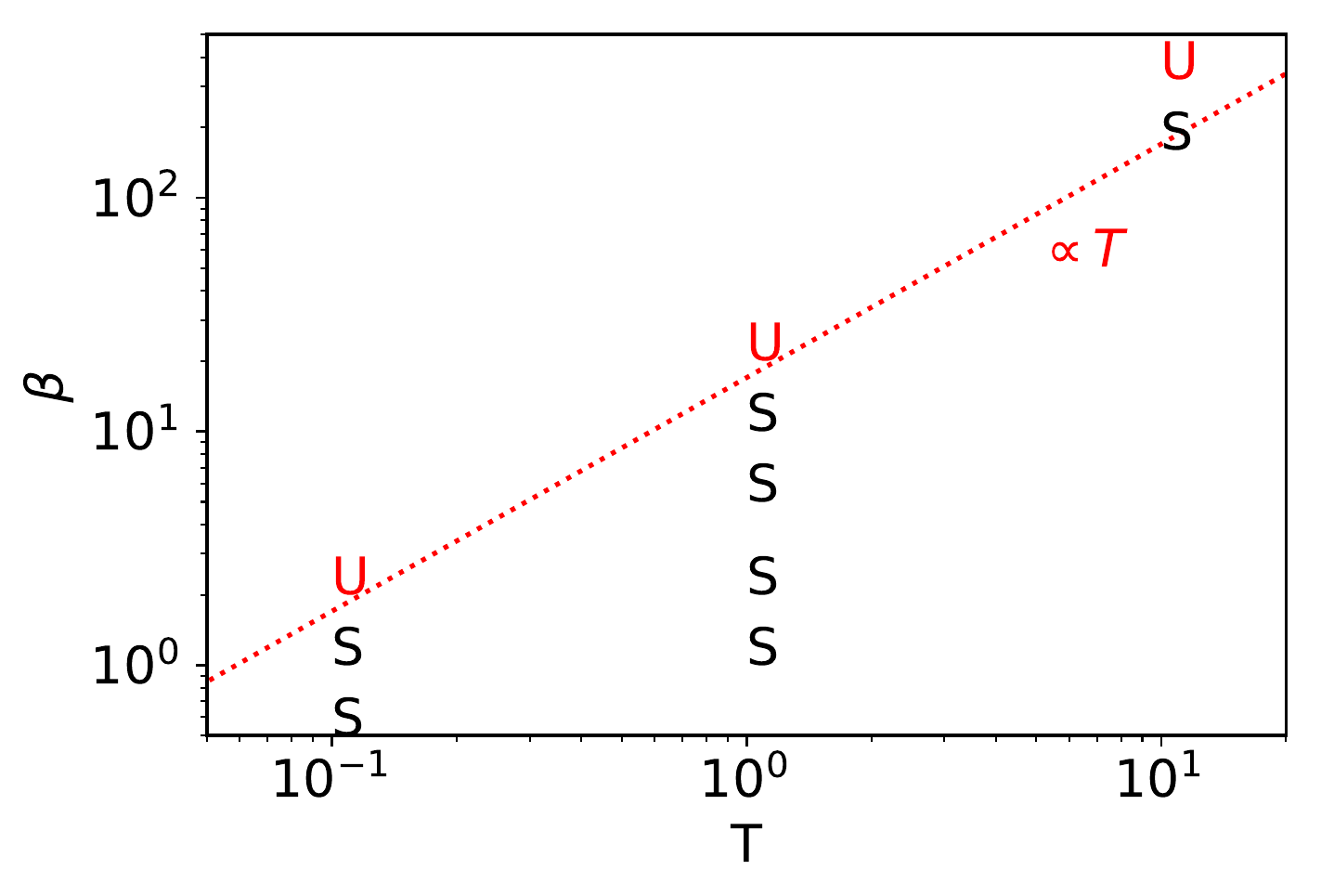}
\includegraphics[width=1.0\columnwidth]{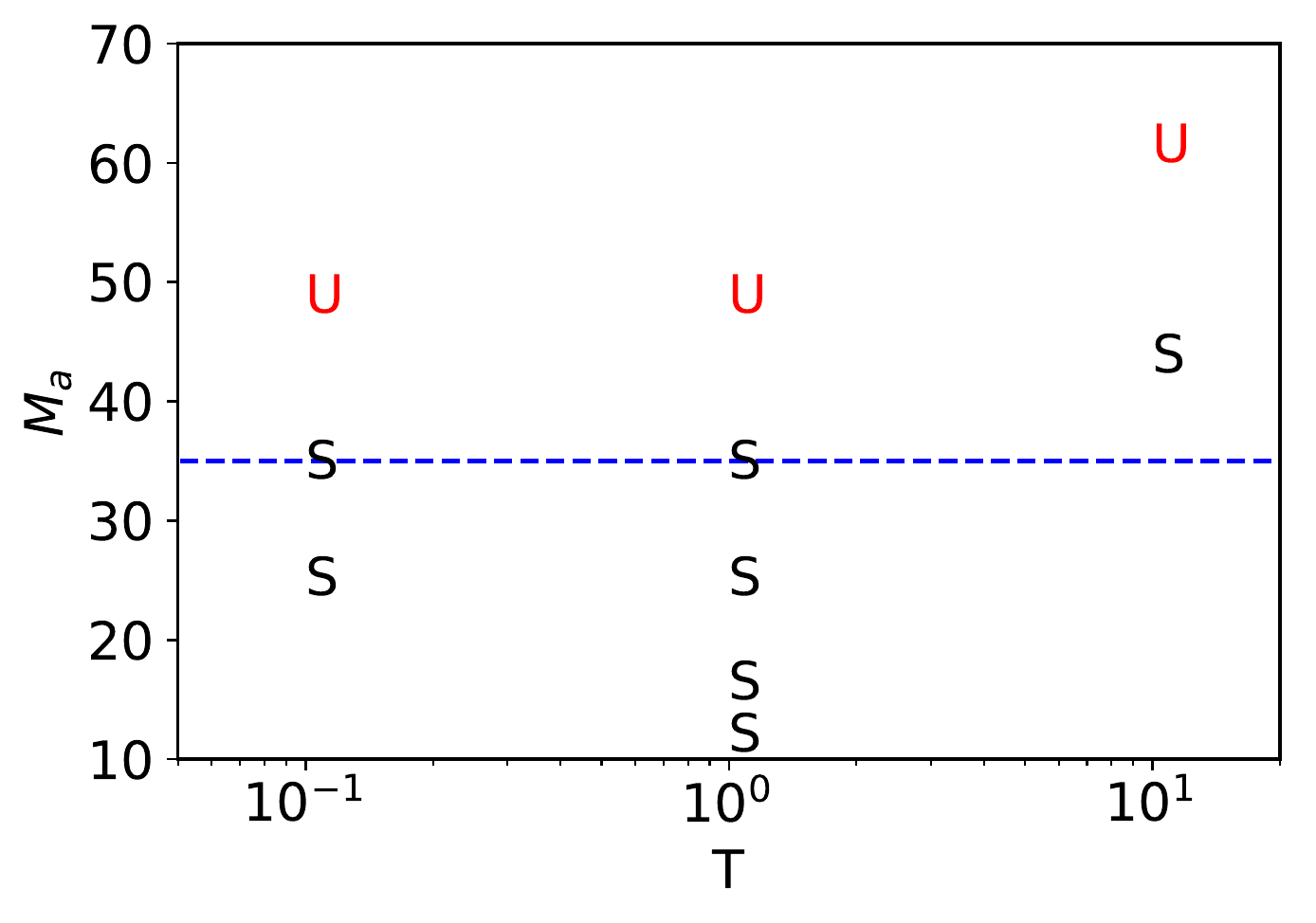}
\caption{Stability indicator for the Newtonian models of series A, B and C in the case of a rotating shell. 
These models have the same  vertical size $\Delta z= 0.01$  of the computational domain and the initial velocity $v_0=15$ but different initial temperature $T$.  The top panels shows the location of stable (S) and unstable (U) models on the $T$-$\beta$ plane and the bottom panel shows the same in the $T$-$M_a$  plane. The blue dashed line at the bottom panel corresponds to the critical Mach number given by equation (\ref{M_crit}). }
\label{Fig:T_NEWT}
\end{figure}
\begin{figure}
\includegraphics[width=1.0\columnwidth]{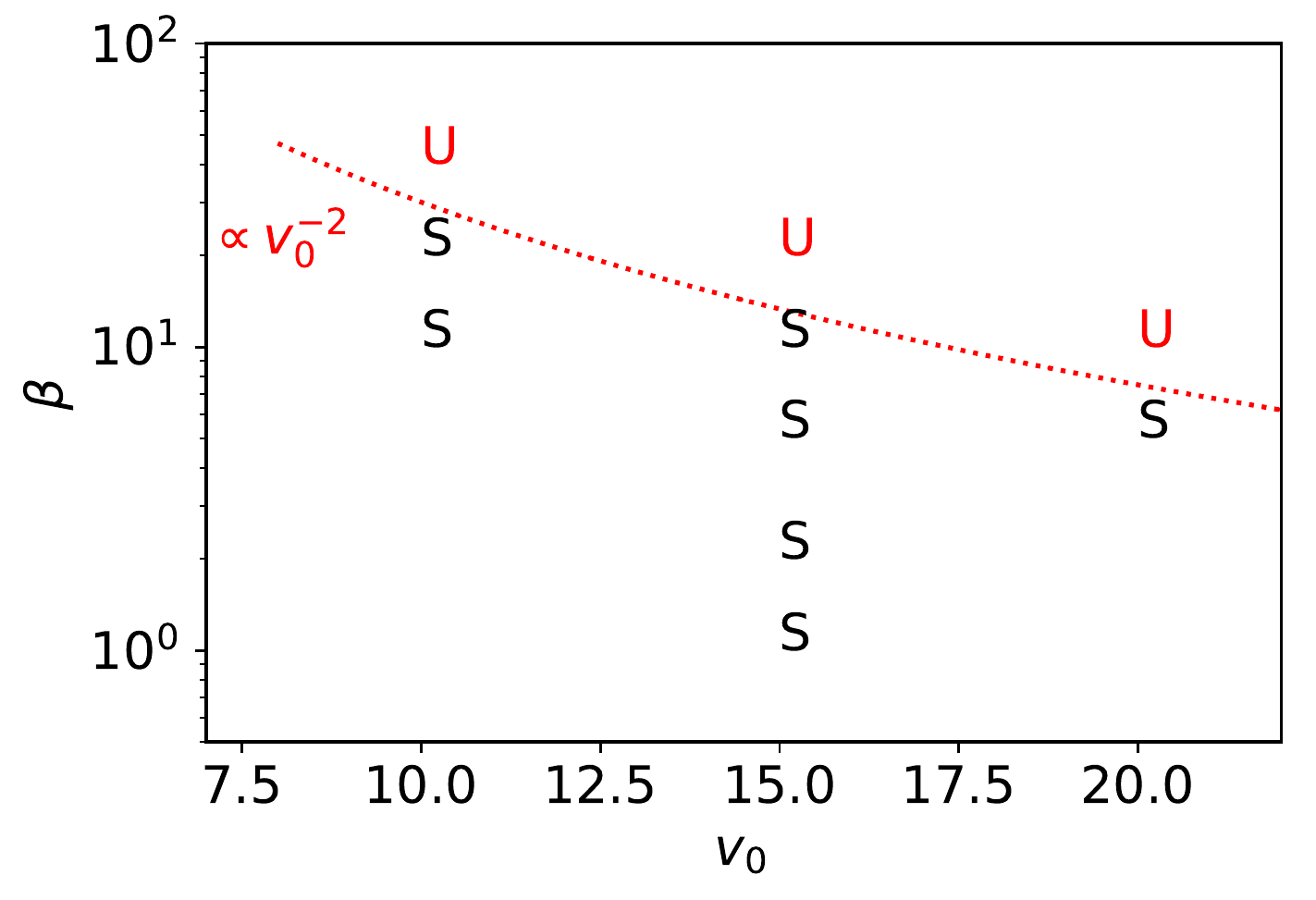}
\includegraphics[width=1.0\columnwidth]{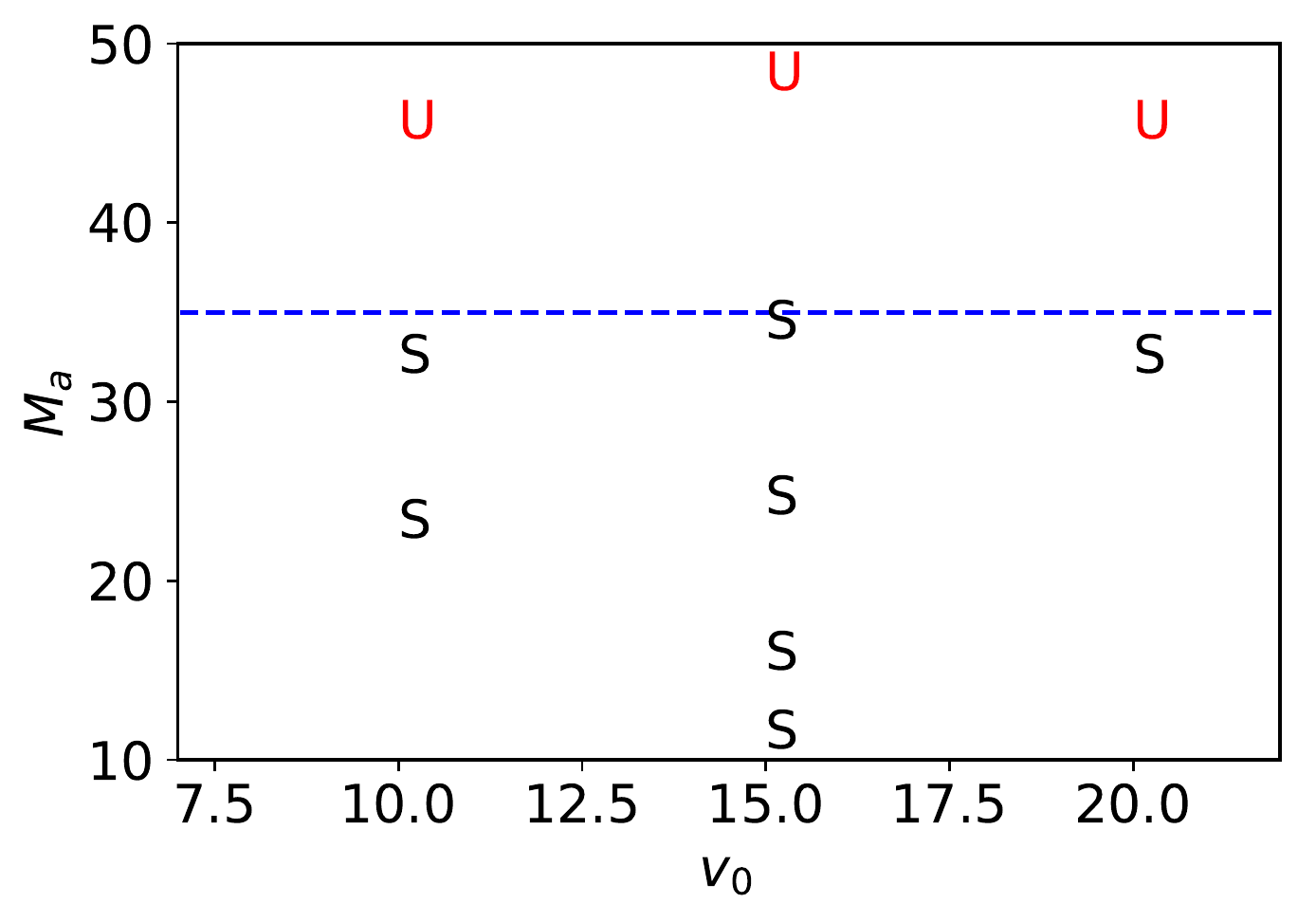}
\caption{
Stability indicator for the Newtonian models of series A, D and E in the case of a rotating shell. 
These models have the same  vertical size $\Delta z= 0.01$  of the computational domain and the initial temperature $T=1$ but different initial velocity $v_0$.  The top panels shows the location of stable (S) and unstable (U) models on the $v_0$-$\beta$ plane and the bottom panel shows the same in the $v_0$-$M_a$  plane. The blue dashed line at the bottom panel corresponds to the critical Mach number given by equation (\ref{M_crit}).}
\label{Fig:v0_NEWT}
\end{figure}

In Table \ref{tab:C} we also give the interface values of the usual Mach number $M_s=v/c_s$, where $c_s=\sqrt{\gamma P/\rho}$ is the sound speed.  As one can see, here we are dealing with at most transonic flows, which is not particularly suitable for the jet problem, where the flow is supersonic. Higher speeds lead to the development of  strong gradients in a thin layer near the interface. This problem is avoided in the rotating shell configuration where the rotating fluid occupies only a thin cylindrical shell. Moreover, as we have pointed out in Sections \ref{Sec:Intro} and \ref{Sec:simulations-overview}, the curved shocked outer layer of a reconfined jet is more like a thin rotating shell rather than a cylinder (In the relativistic simulations we deal with this configuration only.).

\subsection{Rotating shell}
\label{Sec:cs-nc-rs}

 \begin{table*}
	\centering
	\caption{Parameters of Newtonian runs for the case of rotating shell. The parameters are: $v_0$ - the rotation, $\beta$ - the magnetisation parameter, $T$ - the fluid temperature $T$, $a$ the power index of pressure distribution, $M_a$ - the Alfv\'enic Mach number, $M_s$ - the usual Mach number, S/U - the stability indicator, $r\times z$ - the integration domain and  $t_f$ - the total time of the run. }
	\label{tab:B}
	\begin{tabular}{lcccccccccccc} 
		\hline
		Name &$v_0$&$\beta$ & $T$  &$M_a$ &$M_s$ &S/U  & grid & $\Delta r\times \Delta z$ & $t_f$\\
		\hline
		A1&15& 20 & 1 & 48 &12  &U  & $400\times 400$ & $0.005 \times  0.01$ & 1.0\\
		A2&15& 10 & 1  & 34 &12  &S  & $400\times 400$ & $0.005 \times  0.01$ & 1.0\\
		A3&15& 5 & 1 &  24  &12   &S  & $400\times 400$ & $0.005 \times  0.01$ & 1.0\\
		A4&15& 2 & 1 &  15  &12     &S  & $400\times 400$ & $0.005 \times  0.01$ & 1.0\\
		A5&15& 1 & 1 &  11  &12   &S  & $800\times 400$ & $0.01 \times 0.01$ & 1.0\\
		\hline
		B1&15& 320 & 10 &  60 &3.7 &U  & $800\times 160$ 3AMR& $0.025 \times 0.01$ & 1.0\\
		B2&15& 160 & 10 &  42 &3.7 &S  & $800\times 160$ 3AMR& $0.025 \times 0.01$ & 1.0\\
		\hline
		C1&15& 2 & 0.1 &  47 &37  &U  & $400\times 2000$ & $0.001 \times 0.01$ & 1.0\\
		C2&15& 1 & 0.1 &  35  &37  &S  & $400\times 2000$ & $0.001 \times 0.01$ & 1.0\\
		C3&15& 0.5 & 0.1 & 24 &37  &S & $400\times 2000$& $0.001 \times 0.01$  & 1.0\\
		\hline
		D1&10& 40 & 1 &  45 &7.7 &U & $800\times 400$ & $0.01 \times 0.01$ & 1.0\\
		D2&10& 20 & 1 &  32 &7.7 &S &  $800\times 400$ & $0.01 \times 0.01$ & 1.0\\
		D2&10& 10 & 1 &  22 &7.7 &S & $800\times 400$ & $0.01 \times 0.01$  & 1.0\\
		\hline
		E1&20& 10 & 1  & 45 &16 &U & $400\times 1000$ & $0.002 \times 0.01$  & 1.0\\
		E2&20& 5 & 1 &  32  &16 &S & $400\times 1000$ & $0.002 \times 0.01$  & 1.0\\
		\hline
		F1&15& 10 & 1 & 34 &12  &U  & $400\times 800$ & $0.005 \times 0.02$ & 1.0\\
		F2&15& 5 & 1 & 24  &12  &S  & $400\times 800$ & $0.005 \times 0.02$ & 1.0\\
		F3&15& 2 & 1 & 15   &12   &S  & $400\times 800$ & $0.005 \times 0.02$ & 1.0\\
		\hline
		G1&15& 10 & 1 &  34 &12 & U  & $400\times 1600$ & $0.005 \times 0.04$ & 1.0\\
		G2&15& 5 & 1 &  24  &12 &U  & $400\times 1600$ & $0.005 \times 0.04$ & 0.5\\
		G3&15& 2 & 1 & 15  &12 &S  & $400\times 1600$ & $0.005 \times 0.04$ & 1.0\\
		\hline
	\end{tabular}
\end{table*}

For the sake of simplicity, we consider here uniform distributions of the magnetisation parameter $\beta$ and the gas temperature $T=p/\rho$ across the shell in the initial configuration. Hence Eq.~\ref{equilib-n1} reduces to 
\beq
\oder{p}{r} =     A \frac{v_{\hat{\phi}}^2(r)}{r}  p \,,
\label{equil-n2}
\eeq
where $ A=2 /[c_a^2 (1+\beta)]$.  If in addition we impose a uniform velocity profile $v_{\hat{\phi}}=v_0$ then 
Eq.~\ref{equilibr-r} reduces to 
\beq
\oder{p}{r} =  a \frac{p}{r} \,,
\label{eq-equil2}
\eeq
where 
\beq
    a = \frac{2}{1+\beta} M_a^2 = \frac{\beta}{1+\beta}\gamma M_s^2 \,.
\eeq 
Its solution is 
\beq
p = p_{in}  \fracp{r}{r_{in}}^a \,,
\label{eq-sol1}
\eeq 
which for high Mach numbers describes a very rapid increase of the pressure with $r$. In order to avoid the numerical issues related to such a high gradient, we simply set the radial size of the shell to  
\begin{eqnarray}
\Delta r=\frac{1}{a} r_{in}\,,
\label{r-size}
\end{eqnarray}
which is very close to the values obtained earlier based on the jet-shell analogy.
 
If outside of the shell the fluid is at rest then $p=$const. However this leads to a huge velocity jump across one computational cell at the interface, this introduces strong numerical dissipation,  which may even drive strong shock waves from the interface. Such a strong dissipation can be avoided if the jump is replaced with a relatively thin shear layer. In this study we put 
\begin{eqnarray}
    v_{\hat{\phi}} = \left\{ 
    \begin{array}{ccl}
         v_0 & \mbox{if} & r < r_{in} \\
         v_0(r_{b}-r)/(r_{b}-r_{in}) & \mbox{if} & r_{in}<r \leq r_{b} \\
         0 & \mbox{if} & r > r_{b}
    \end{array} \right.
\end{eqnarray}
where $r_{b}=r_{in}+0.1\Delta r$. Hence for $r<r_{in}$ the gas pressure is still given by Eq.~\ref{eq-sol1}. As one can easily verify by integrating Eq.~\ref{equil-n2}, for $r>r_{in}$ the pressure distribution is also given by an analytic function but it is a bit cumbersome and hence not worthy of being presented.

\begin{table*}
	\centering
	\caption{Parameters of the relativistic runs.  $\Gamma$ - the initial Lorentz factor, $\sigma=b^2/(b^2+4\pi w)$ - the relativistic magnetisation parameter, $T=p/\rho c^2$ - the fluid temperature, $\beta=p/p_m$, - the Newtonian magnetisation parameter, $M_s$ - the Mach number with respect to the sound speed, $M_a$ - the Mach number with respect to the  Alfv\'en speed, S/U - the stability indicator, $\Delta r \times \Delta z$ - the size of the integration domain,  $t_f$ - the total time of the run. }
	\label{tab:A}
	\begin{tabular}{lccccccccccc} 
		\hline
		Name &$\Gamma_0$&$\sigma$ & $T$ & $\beta$ & $M_s$ &$M_a$& S/U  & grid & $\Delta r \times \Delta z$ & $t_f$\\
		\hline
		A1 &5& 0.01 & 0.1 & 14. & 15.  & $49.$&U & $400^2$ & $0.005 \times 0.01$ & 2.0\\
		A2 &5& 0.02 & 0.1 & 7.1 &  15. & $35.$& U& $400^2$ & $0.005 \times 0.01$ & 2.0\\
    		A3&5& 0.04 & 0.1 & 3.6 &   15. & $25.$& S & $400^2$ & $0.005 \times 0.01$ & 8.0\\
		\hline
		B1 &5& 0.04 & 0.1 & 3.6 &  15.  & $25.$& U & $400\times 800$ & $0.005 \times 0.02$ & 2.3\\
		B2 &5& 0.08 & 0.1 & 1.8 &  15.  & $17.$& S & $400\times 800$ & $0.005 \times 0.02$ & 8.0\\
		\hline
		C1 &5& 0.04 & 0.1 & 3.6 &  15.   & $25.$& U & $400\times 1600$ & $0.005 \times 0.04$ & 2.0\\
		C2 &5& 0.08 & 0.1 & 1.8 &  15.  & $17.$& S & $400\times 1600$ & $0.005 \times 0.04$ & 8.0\\
		\hline
		D1 &5& 0.01 & 1.0 & 40. & 8.1   & $49.$& U & $400^2$ & $0.005 \times 0.01$ & 1.2\\
		D2 &5& 0.02 & 1.0 & 20. & 8.1  & $35.$& S & $400^2$ & $0.005 \times 0.01$ & 8.0\\
		\hline
		E1&3& 0.01 & 0.1 & 14. &  8.7  & $28$& U & $800^2$ & $0.01 \times 0.02$ & 2.7\\
		E2&3& 0.02 & 0.1 & 7.1 &   8.7 & $20$ & S & $800^2$ & $0.01 \times 0.02$ & 8.0\\
		E3&3& 0.04 & 0.1 & 3.6 &   8.7 & $14$& S & $800^2$ & $0.01 \times 0.02$ & 8.0\\
		\hline
		F1&5& 0.005 & 10. & 98. &  7.1  & $69$& U & $400^2$ & $0.005 \times 0.01$ & 1.0\\
		F2&5& 0.01 & 10. & 49. &   7.1 & $49$& U & $400^2$ & $0.005 \times 0.01$ & 1.0\\
		F3&5& 0.02 & 10. & 24. &   7.1 & $35$& S & $400^2$ & $0.005 \times 0.01$ & 8.0 \\ 
		\hline
		G1&5& 0.005 & 100. & 100. & 6.9   & $69$& U & $400^2$ & $0.005 \times 0.01$ & 0.4\\
		G2&5& 0.01 & 100. & 50. &   6.9 & $49$ & U & $400^2$ & $0.005 \times 0.01$ & 0.5\\
		G3&5& 0.02 & 100. & 25. &   6.9 & $34$& S & $400^2$ & $0.005 \times 0.01$ & 8.0 \\ 
		\hline 
		H1&10& 0.01 & 0.1 & 14. & 31.   & $99$& U & $400\times 4000$ & $0.001 \times 0.02$ & 0.2\\
		H2&10& 0.02 & 0.1 & 7.1 & 31.   & $70$& U & $400\times 4000$ & $0.001 \times 0.02$ & 0.4\\
		H3&10& 0.04 & 0.1 & 3.6 & 31.   & $50$ & U & $400\times 4000$ & $0.001 \times 0.02$ & 1.5\\
		H4&10& 0.08 & 0.1 & 1.8 & 31.   & $35$& S & $200\times 2000$ 2AMR & $0.001 \times 0.02$ & 8.0\\
		H5&10& 0.16 & 0.1 & 0.89 & 31.   & $25$ & S & $200\times 2000$ 2AMR& $0.001 \times 0.02$ & 8.0\\
		H6&10& 0.32 & 0.1 & 0.45 & 31.   & $18$ &S & $200\times 1000$ 2AMR& $0.001 \times 0.02$ & 8.0\\
		\hline 
		I1&5& 0.01 & 0.05 & 8.3 &  20.  & $49$& U & $160\times 400$ 3AMR& $0.002 \times 0.01$ & 8.0\\
		I2&5& 0.02 & 0.05 &  4.2&  20.  & $35$&S & $160\times 400$  3AMR& $0.002 \times 0.01$ & 8.0\\
		I3&5& 0.04 & 0.05 &  2.1&  20.  & $24$& S & $160\times 400$ 3AMR& $0.002 \times 0.01$ & 8.0\\
		\hline 
		J1&1.5& 0.001 & 0.1 & 140. & 3.4   & $35$& U & $800\times 160$ 3AMR& $0.05 \times 0.02$ & 1.2\\
		J2&1.5& 0.002 & 0.1 & 71. &  3.4   & $25$& S& $800\times 160$ 3AMR& $0.05 \times 0.02$ & 8.0\\
		J3&1.5& 0.004 & 0.1 & 36. &  3.4  & $18$& S & $800\times 160$ 3AMR & $0.05 \times 0.02$ & 8.0\\
		\hline
		K1 &7& 0.02 & 0.1 & 7.1 &  21.  & $50$& U & $400\times 2000$ & $0.002 \times 0.02$ & 0.6\\
		K2 &7& 0.04 & 0.1 & 3.6 &   21.  & $35$& U & $400\times 2000$ & $0.002 \times 0.02$ &1.0\\
		K3 &7& 0.08 & 0.1 & 1.8 &   21.  & $25$& S & $400\times 2000$ & $0.002 \times 0.02$ & 8.0\\
		\hline
	\end{tabular}
\end{table*}

Our approach in the simulations is to fix the fluid temperature (sound speed) and the shell velocity, and then to vary the magnetisation parameter $\beta$ until we observe a transition from stable to unstable behaviour. In Table~\ref{tab:B} the runs of such a series are identified by the first letter of their name (e.g. A) and individual runs within a series are identifies by the number following the letter (e.g.  A1, A2, etc).  Then we change the fixed parameters and repeat the process.  The ultimate goal is to identify the critical parameter (or parameters) defining the transition.  The results are presented in Table~\ref{tab:B}.    

Series A, B and C differ only by the shell temperature. Their results are illustrated in Figure~\ref{Fig:T_NEWT}.   
One can see that the critical magnetisation depends on the fluid temperature and that the results are consistent with 
$\beta^* \propto T$.   Along this line the Alfv\'en speed $c_a$, and hence the Alfv\'enic Mach number $M_a$, is constant. 
This is consistent with the lower panel of this figure which shows the data in the $M_a$-$T$ plane.  Thus, the results are 
suggestive of  $M_a$ being the most important parameter determining the transition to instability.        

Series A, D and E differ only by the shell velocity. Their results are illustrated in Figure~\ref{Fig:v0_NEWT}. One can see that 
the critical magnetisation parameter depends on the shell velocity approximately as $\beta^* \propto v_0^{-2}$. Along this line 
$M_a$ is constant again, providing further support to the above conclusion.

\section{Computer simulations. The relativistic case}
\label{Sec:cs-rc}

As we have shown in Sec.~\ref{Subsec:HR}, in the steady state the flow must satisfy the equation 

\beq
\frac{d(p+p_m)}{dr}= \frac{(w+2p_m)u_{\hat{\phi}}^2}{r}\,.
\label{equilibr-r}
\eeq
This equation involves four unknown functions and hence allows a rich family of solutions. We chose a configuration where the key magnetisation parameters $\beta=p/p_m$ and $\sigma= 2p_m/w$ are constant throughout the domain. 
These two parameters determine the gas temperature $T=p/\rho c^2$ via  
\begin{eqnarray}
\sigma=\frac{2}{\beta} \frac{T}{1+\kappa T} \,,
\end{eqnarray}
where $\kappa=\gamma/(\gamma-1)$ and hence $T$ is also constant. Under these assumptions, Eq.~\ref{equilibr-r} reduces to 
\beq
\oder{p}{r} =     A \frac{u_{\hat{\phi}}^2(r)}{r}  p \,,
\label{eq-equil1}
\eeq
where 
$ A=2 (1+\sigma) /((1+\beta) \sigma) $.
The standard jet model assumes that its velocity (Lorentz factor) is uniform over the jet cross-section and so we assume that  $u_{\hat{\phi}}$ is also constant and equals to $u_0$ as well. This leads to the equilibrium equation 
\beq
\oder{p}{r} =  a \frac{p}{r} \,,
\label{eq-equil-r2}
\eeq 
and hence the pressure distribution  
\beq
p = p_{in}  \fracp{r}{r_{in}}^a \,.
\label{eq-sol-r1}
\eeq 
Like in the Newtonian case, the power index  
\beq
    a = (1+\sigma) \frac{2}{1+\beta} M_a^2 
\eeq 
can be quite high when the Alfv\'enic Mach number $M_a =\Gamma v_{\hat{\phi}}/\Gamma_a c_a \gg 1$, which may result in a strong pressure variation, and in order to avoid this we set the radial size of the shell to  
\begin{eqnarray}
\Delta r=\frac{1}{a} r_{in}\,,
\label{r-size-r}
\end{eqnarray}

Like in the Newtonian case, we introduce a shear layer between the shell and the external fluid in order to reduce the numerical dissipation.  The resulting velocity profile is 
\begin{eqnarray}
    u_{\hat{\phi}} = \left\{ 
    \begin{array}{ccl}
         u_0 & \mbox{if} & r < r_{in} \\
         u_0(r_{bl}-r)/(r_{bl}-r_{in}) & \mbox{if} & r_{in}<r \leq r_{bl} \\
         0 & \mbox{if} & r > r_{bl}
    \end{array} \right. \,,
\end{eqnarray}
where $r_{bl}=r_{in}+0.1\Delta r$.  For $r\le r_{in}$ the gas pressure distribution is given by Eq.~\ref{eq-sol-r1}, for $r_{in}<r<r_{bl}$ it is found via integrating Eq.~\ref{eq-equil1} and it is  constant for $r>r_{bl}$. Given the uniform magnetisation, the gas pressure is continuous both at $r=r_{bl}$ and $r_{in}$.

Here we use almost the same strategy as in the Newtonian simulations of rotating shells -- we fix the fluid temperature (sound speed) and the shell Lorentz factor, and then vary the magnetisation parameter $\sigma$ (Alfv\'en speed) until we observe a transition from stable to unstable behaviour. In Table~\ref{tab:B} the runs of such a series are identified by the first letter of their name (e.g. A) and individual runs within a series are identifies by the number following the letter (e.g.  A1, A2, etc).  Then we change the fixed parameters and repeat the process.  The ultimate goal is to identify the critical parameter (or parameters) defining the transition.  The results are presented in Table~\ref{tab:A}.

\begin{figure}
\includegraphics[width=0.95\columnwidth]{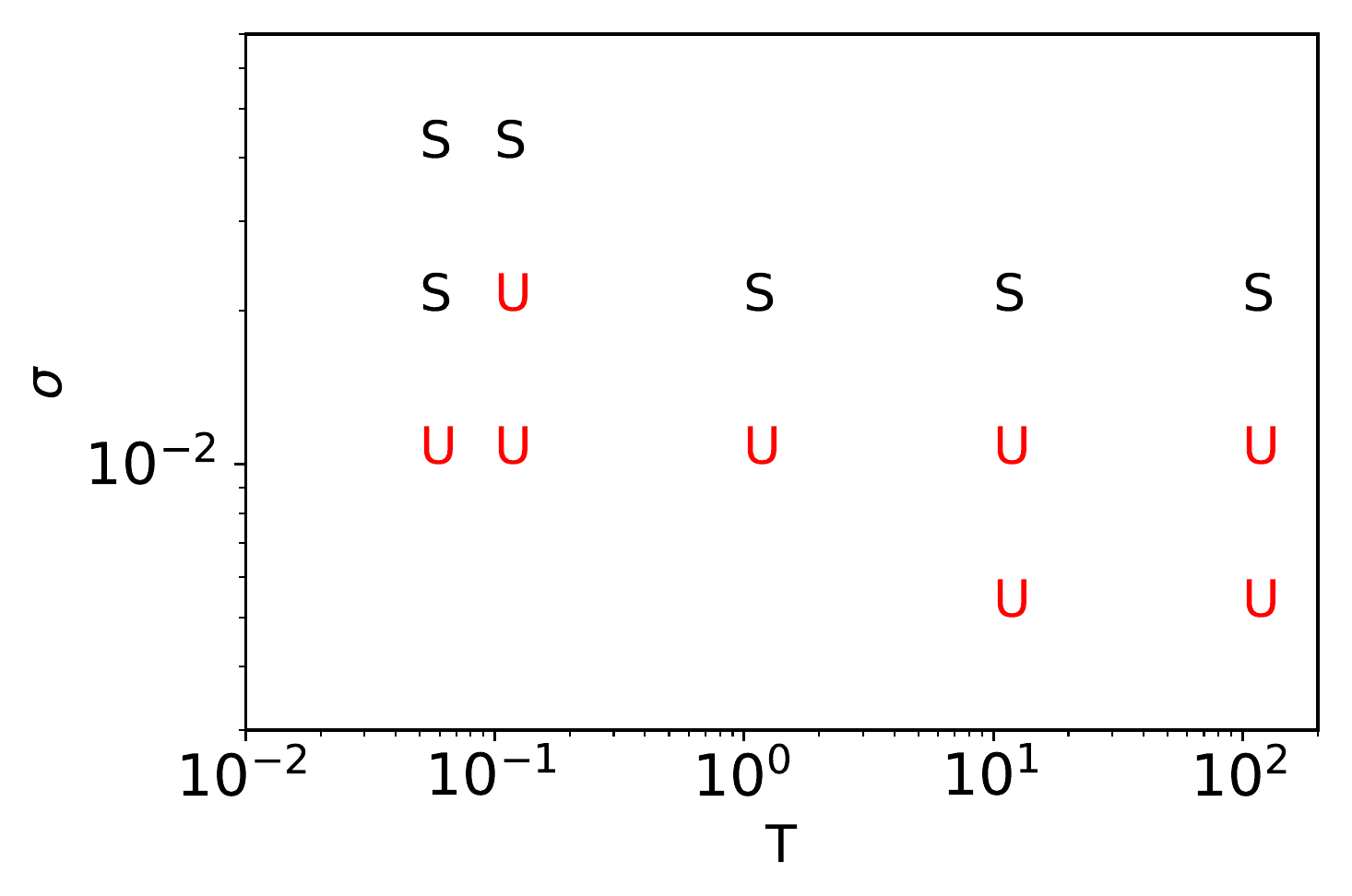}
\includegraphics[width=0.95\columnwidth]{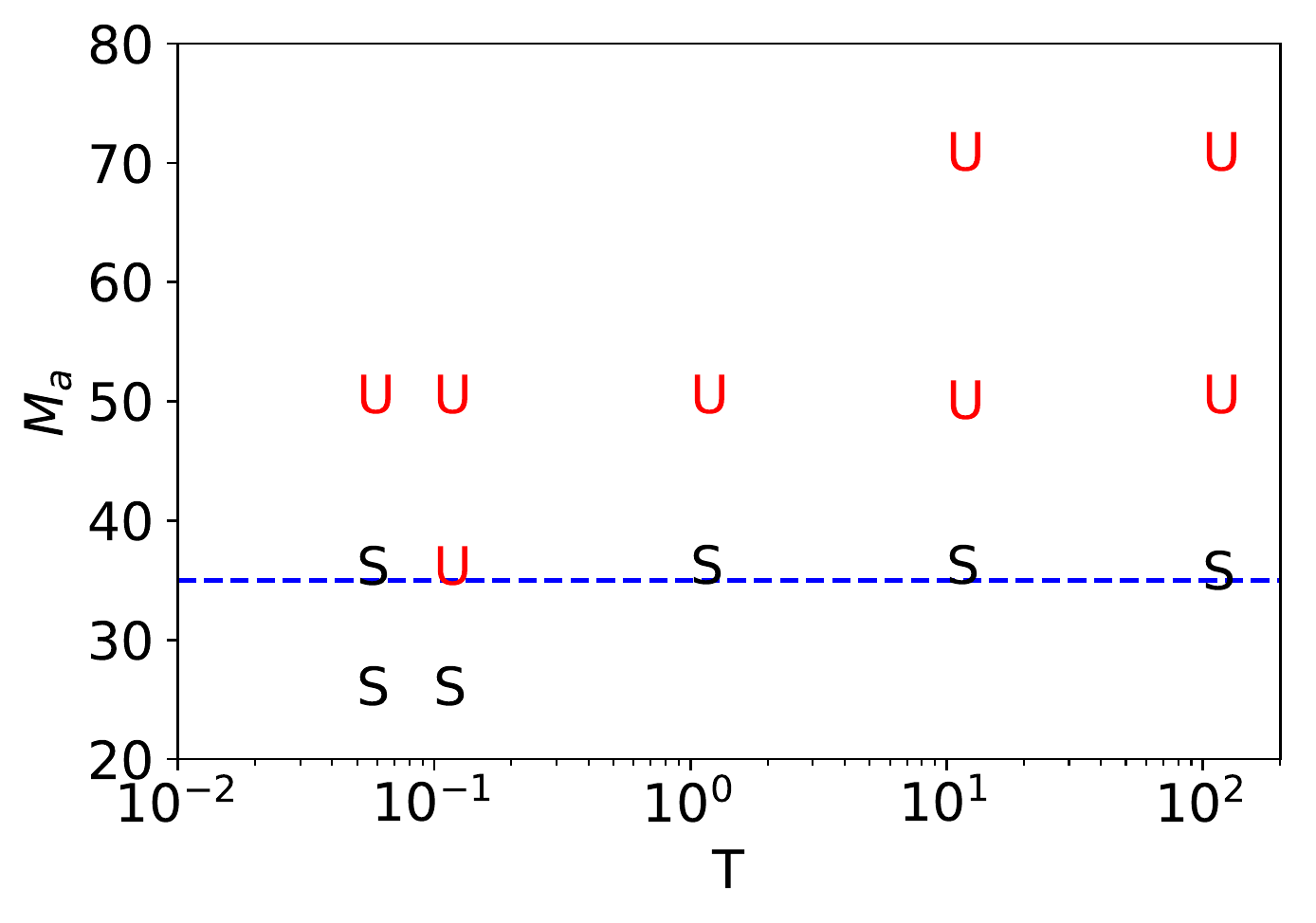}
\caption{ Stability indicator for the relativistic models of series A, D, F, G and I, which have the same vertical size of the computational domain $\Delta z= 0.01$  and the initial Lorentz factor $\Gamma _0=15$ but different initial temperature $T$.  The top panels shows the location of stable (S) and unstable (U) models in the $(T,\sigma)$ plane and the bottom panel shows the same in the $(T,M_a)$  plane. The blue dashed line at the bottom panel corresponds to the critical Mach number given by equation (\ref{M_crit}). }
\label{Fig:G05}
\end{figure}

Series A, D, F, G and I have the same $\Gamma=5$, the same $\Delta z=0.01$ but differ by the shell temperature, covering $T\in[0.05,100]$. Their results strongly suggest that the temperature has little effect, if any, on the onset of the instability (see Figure~\ref{Fig:G05}). Indeed, in all these models the transition occurs within the narrow range of the magnetisation $0.01 < \sigma < 0.02$. However, in the $M_a$-$T$ plane we see a similar picture, with the transition occurring within the narrow range of the Mach number  $35<M_a<49$, making $M_a$ another candidate. Obviously, this is simply because $\sigma$ uniquely determines the Alfv\'en speed and with fixed $\Gamma$ the Mach number as well. 
To resolve the degeneracy we need to vary $\Gamma$.   
 
\begin{figure}
\includegraphics[width=0.95\columnwidth]{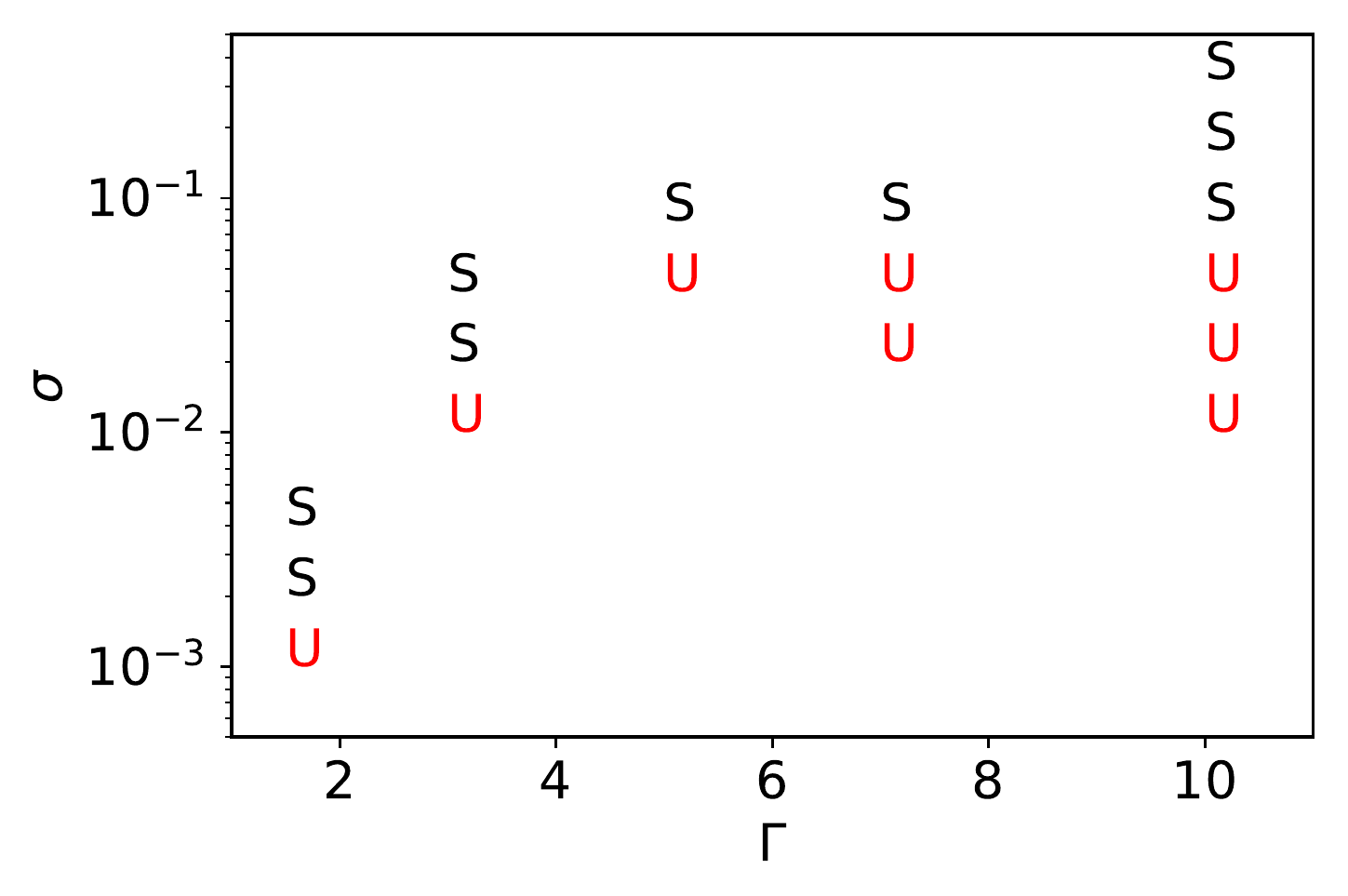}
\includegraphics[width=0.95\columnwidth]{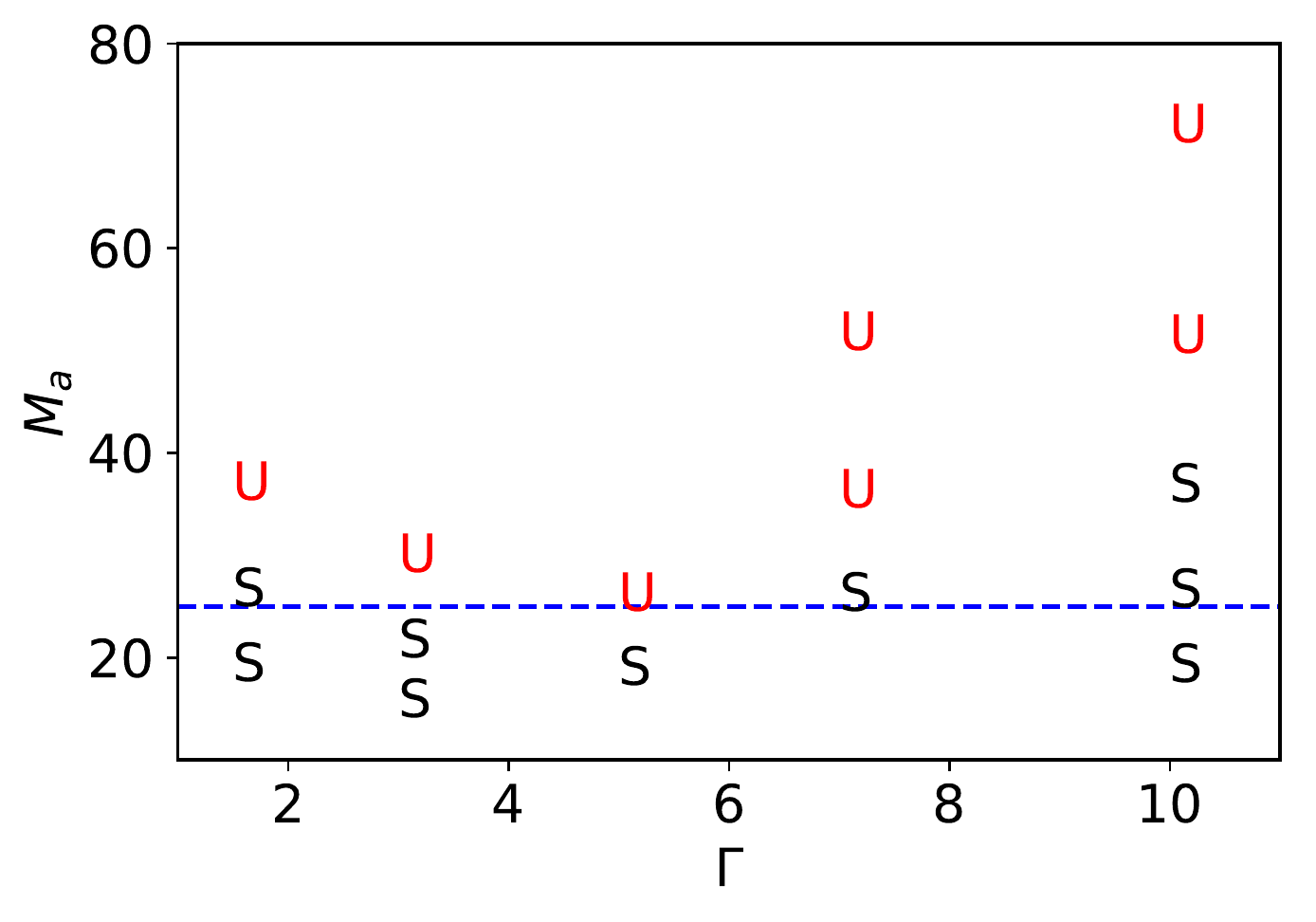}
\caption{Stability indicator for the relativistic models of series B, E, H, J and K, which have the same vertical size of the computational domain $\Delta z= 0.02$  and the initial temperature  $T=0.1$ but different initial Lorentz factor $\Gamma$.  The top panels shows the location of stable (S) and unstable (U) models in the $(\Gamma,\sigma)$ plane and the bottom panel shows the same in the $(\Gamma,M_a)$ plane. The blue dashed line at the bottom panel corresponds to the critical Mach number given by equation (\ref{M_crit}). }
\label{Fig:T01}
\end{figure}

Series B, E, H, J and K have the same $T=0.1$, the same $\Delta z=0.02$ and differ by the shell $\Gamma$. Their results are illustrated in Figure~\ref{Fig:T01}. One can see that the critical value of $\sigma$ varies rapidly for $\Gamma<5$, 
with the overall variation about 1.5 orders of magnitude. This shows that the transition to instability is not uniquely determined by the magnetisation parameter $\sigma$. In the $M_a$-$\Gamma$ plane the division line is not a straight horizontal either but the deviations from it are not so dramatic, with the critical $M_a$ varying by no more than the factor of 2.5.

\section{Discussion}
\label{Discussion}

\subsection{Testing the heuristic criterion against the simulation results}

\subsubsection{Newtonian case. Rotating cylinder} 

When the outer fluid of this configuration is not rotating ($v_{\phi,2}=0$, models of the group A), the simulations suggest that the critical parameter is the Alv\'enic Mach number  $M_{a,1} = v_{\hat{\phi,1}}/c_{a,1}$ of the inner fluid at the interface.  In this case, the non-relativistic expression (\ref{lambda-cr-n}) for the critical wavelength reduces to   
\beq  
       \lambda_c = \frac{B^2}{ \rho v_{\hat{\phi},1}^2 } r_{in} = 4\pi r_{in} \frac{1}{M_{a,1}^2} \,, 
      \label{lambda-cr-A}
\eeq  
which is indeed consistent with the finding.  Obviously, the instability can develop only when the critical wavelength becomes smaller than the $z$ size of the computational domain. Hence this condition sets the critical Mach number 
 \beq
    {M_{a,1}^{*}} = \sqrt{4\pi \frac{r_{in}}{\Delta z}} \,.  
\eeq     
For $r_{in}=1$ and $\Delta z=2$ this yields $M_{a,1}^{*}\approx 2.5$, whereas in the simulations the transition from stable to unstable regime occurs for $1.4<M^{*}_{a,1}<2.8$ (see Table~\ref{tab:C}).  Hence we have got a good quantitative agreement between the theory and the simulations.

When both fluids are rotating, the critical wavelength can be written as  
\beq  
       \lambda_c =  4\pi r_{in} \frac{1}{M_{a,1}^2 - M_{a,2}^2} \,.
      \label{lambda-cr-BC}
\eeq  
Hence in this case the transition between stable and unstable regimes is governed by the parameter
$$
   \delta M_a = (M_{a,1}^2 - M_{a,2}^2 )^{1/2} \,,
$$
with the critical value being equal to
\beq
   \delta M_a^{*}= \sqrt{4\pi \frac{r_{in}}{\Delta z}} \,.  
\eeq     
For $r_{in}=1$ and $\Delta z=2$ this yields the $\delta M_{a}^{*}\approx 2.5$. In the simulation, the transition occurs for $ 2.0 < \delta M_a^{*}<4.9$ in the case of group B and  $ 1.8 < \delta M_a^{*}<2.6$ in the case of group C. Thus for all the three groups of the Newtonian models the simulations results agree with the heuristic instability criterion (\ref{lambda-cr-n}). 

\begin{figure}
\includegraphics[width=1.0\columnwidth]{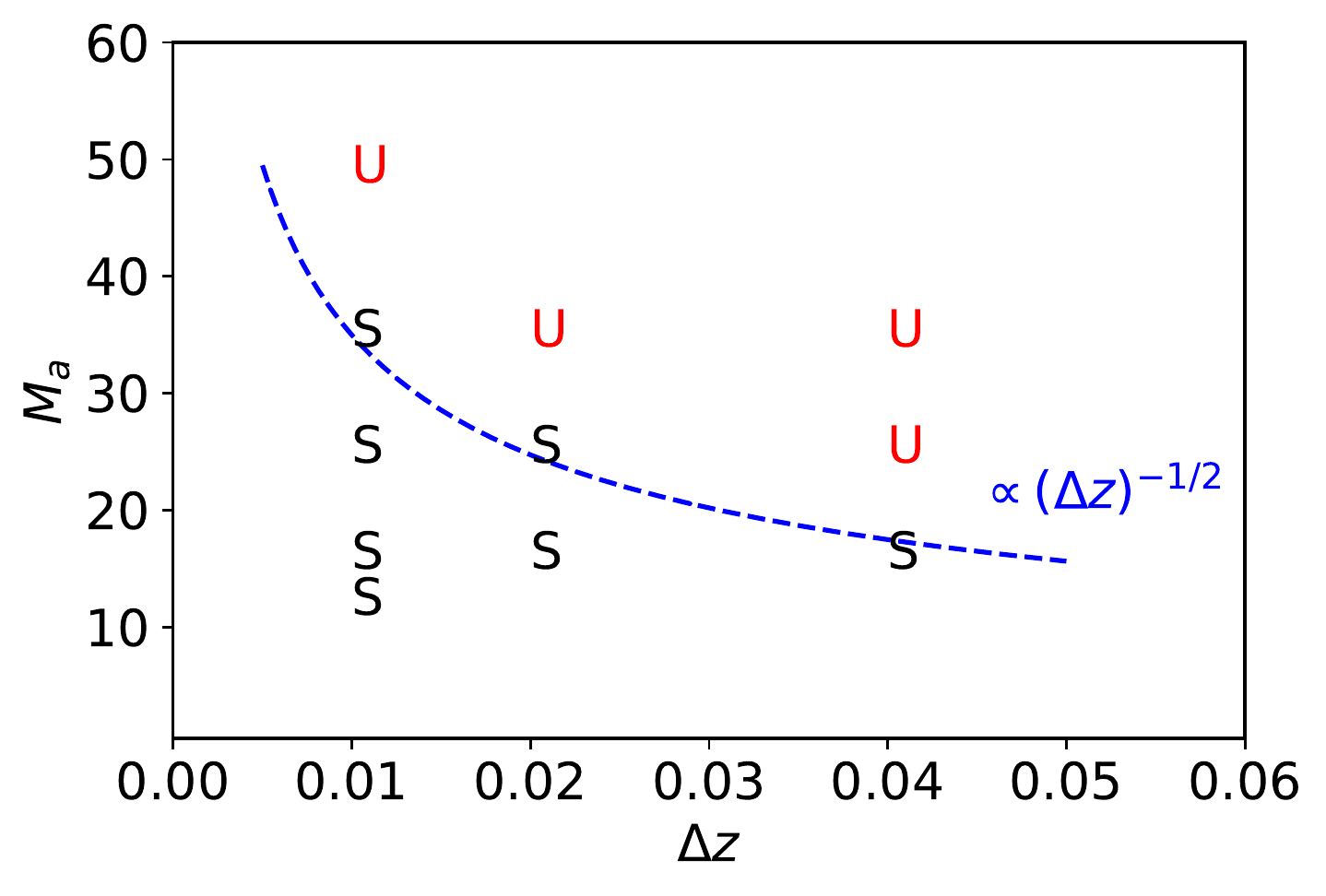}
\caption{
Stability indicator for the Newtonian models of series A, F and G in the case of a rotating shell. 
These models have the same velocity $v_0= 15$ and the initial temperature $T=1$ but different vertical size of the computational domain $\Delta z$.  The location of stable (S) and unstable (U) models on the $\Delta z $-$M_a$ plane is shown. The blue dashed line corresponds to the critical Mach number given by equation (\ref{M_crit})..}
\label{Fig:Dz_Ma_Newt}
\end{figure}

\subsubsection{Newtonian case. Rotating shell} 

As far as the interface is concerned, this case is no different from the group A models of the rotating-cylinder configuration. 
Hence the heuristic instability criterion yeilds the same expression for the critical Alv\'enic Mach number  
 \beq
    M_{a}^{*} = \sqrt{4\pi \frac{r_{in}}{\Delta z} }\,.  
    \label{M_crit}
\eeq     
For $r_{in}=1$ and $\Delta z=0.01$ we have $M_{a}^{*}\approx 35$, which agrees quite well with the data of Table (\ref{tab:B}). We note that here we are dealing with much higher Alv\'enic Mach numbers then in the rotating-cylinder case, which suggests that the criterion is quite robust.

Series A, F and G differ only by $z$ size of the computational domain, allowing modes with progressively longer wavelenghts. 
The results show that higher $\Delta z$ corresponds to lower critical Alfv\'en Mach number. This is fully consistent with Eq.~\ref{M_crit},
which predicts $M_a^{*} \propto 1/\sqrt{\Delta z}$, as illustrated in Figure~\ref{Fig:Dz_Ma_Newt}.
This is fully consistent with the expectation that stabilisation of modes with longer wavelengths requires stronger magnetic field. 

\begin{figure}
\includegraphics[width=0.95\columnwidth]{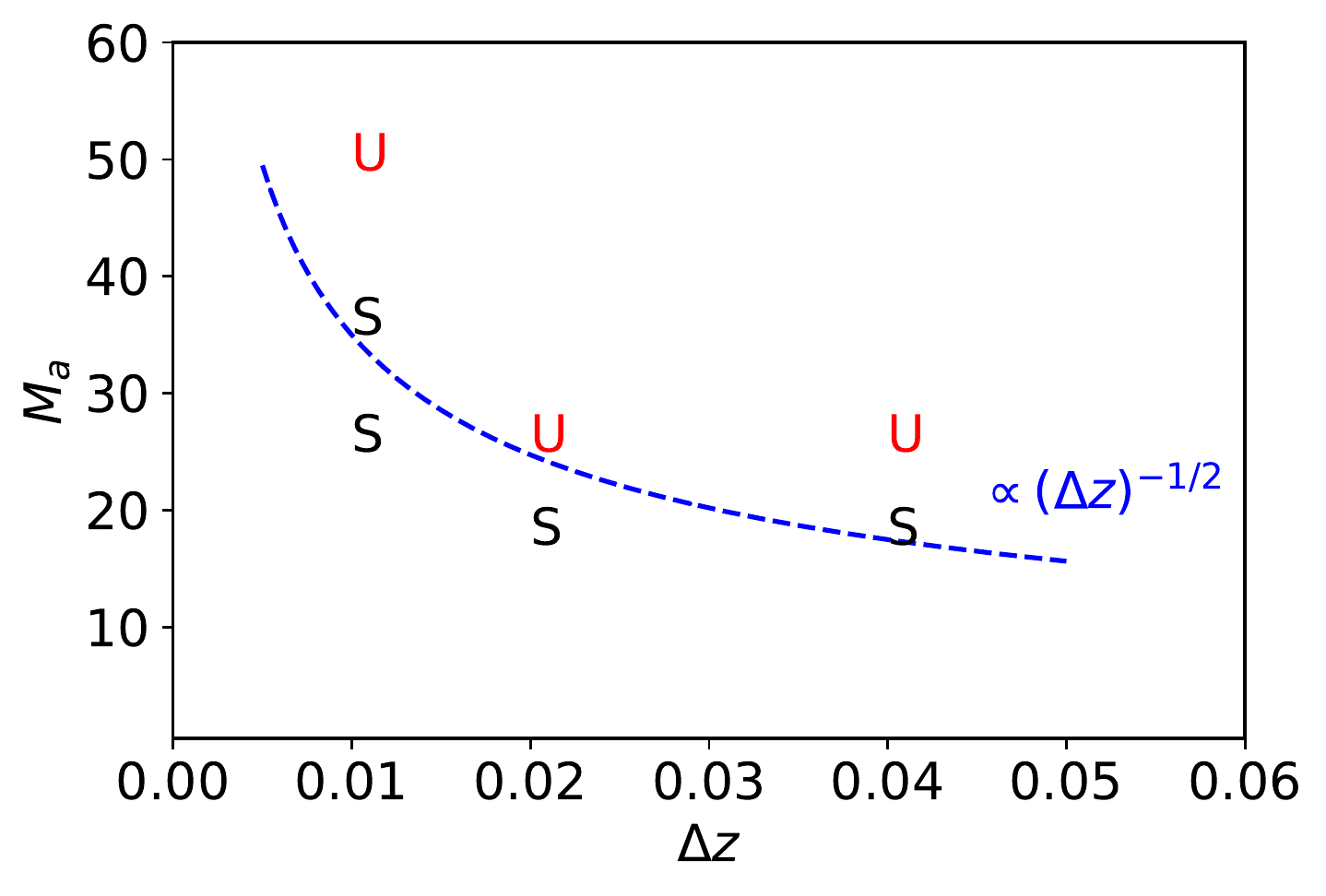}
\caption{Stability indicator for the relativistic models of series A, B and C which have the same Lorentz factor $\Gamma=5$, initial temperature  $T=0.1$ but different on the vertical size of the computation domain $\Delta z$. The location of stable (S) and unstable (U) models is in the $(\Delta z, M_a)$ plane is shown. The blue dashed line corresponds to the critical Mach number given by equation (\ref{M_crit}). }
\label{Fig:Dz_Ma_Rel}
\end{figure}

\subsubsection{Relativistic case. Rotating shell} 

For $v_{\phi,1}=0$, the relativistic criticality condition (\ref{lambda-cr-r}) reduces to 
\beq  
       \lambda_c = 4\pi \frac{b^2}{ (4\pi w + b^2) u_{\hat{\phi},1}^2 } r_{in} =  4\pi r_{in} \frac{1}{M_a^2}\frac{1}{1+\sigma} \,,
      \label{lambda-cr-r1}
\eeq  
where $\sigma=b^2/4\pi w$ and $M_a=(\Gamma v)/(\Gamma_a c_a)$ is the relativistic Mach number with respect to the Alfv\'en speed $c_a^2=b^2/(4\pi w + b^2)$. For small $\sigma$, which is the case in our simulations, this expression is the same as the corresponding Newtonian equation (\ref{lambda-cr-A}) and hence the critical Mach number is still given by the Newtonian result (\ref{M_crit}). 
For  $\Delta z=0.01$ and $r_{in}=1$, equation (\ref{M_crit}) yields $M_{a}^*\approx 35$. In figure~\ref{Fig:G05} this prediction is compared with the results for the numerical models with $\Gamma_0=5$ and $\Delta z=0.01$, used to probe the dependence of the instability on the plasma temperature. One can see that they are consistent,  with the theoretical value being probably just a little bit lower than the data.  

For  $\Delta z=0.02$ and $r_{in}=1$, equation  (\ref{M_crit}) yields $M_{a}^*\approx 25$. In figure~\ref{Fig:T01} this prediction is compared with the results for the models used to probe the instability dependence on the Lorentz factor.  One can see that they agree very well for $3\lesssim \Gamma \lesssim 6$ but some deviations seems to emerge for low and high Lorentz factors. The deviation for $\Gamma<3$ is unlikely to be significant as the in the Newtonian case the theory and the simulations agree quite well.   In contrast, the deviation for $\Gamma>7$ could indicate that in our analysis we have missed some of the relativistic effects. In order to verify that this is a reliable result and not a numerical artefact,  we repeated the relativistic runs J2, K3 and H4 with the doubled resolution, via adding an extra adaptive mesh refinement level, and found no difference in their stability properties. 

At high $\Gamma$ the growth rate of the instability can be reduced due to the time dilation effect. In principle, this could result in erroneous  classification of a model as stable if the simulations did not run for sufficiently long time. However, this is unlikely to be the correct explanation as the time dilation effect is equally pronounced for all the models of the group H, both stable and unstable. In order to be absolutely sure that this is the case, we have run the stable models for much longer compared to the unstable ones with the same $\Gamma$ and check that this has no effect on the outcome.  

The results of the relativistic simulations seem to agree with the prediction  $M^*_a \propto 1/\sqrt{\Delta z}$ of the heuristic stability criterion (see Figure \ref{Fig:Dz_Ma_Rel}, which based on the data for the A, B and C models).

\subsection{The MRI modes}

In the Newtonian framework, the distinction between CFI and MRI modes is best exposed in the simplified 
model proposed by \citet{Balbus-03}. It leads to the dispersion relation 

\be 
      \omega^4 -(2K^2+\kappa^2) \omega^2 + K^2(K^2+r\oder{\Omega^2}{r}) = 0 \,,
      \label{disp-eq}
\ee 
where 
\be
   \kappa^2 = 4\Omega^2 + r \oder{\Omega^2}{r} \,,
\ee
$K^2 = k^2 c_a^2 \cos^2\alpha$, $k=2\pi/\lambda$ is the wavenumber of the perturbation and $\alpha$ is the angle between
the wavevector and the magnetic field.  

In the limit of vanishing magnetic field, $K=0$ and equation (\ref{disp-eq}) reduces to 
\be
     \omega^2(\omega^2-\kappa^2) = 0 \,.  
\ee 
Provided $\kappa^2>0$, which can also be written as the Rayleigh stability criterion
\be
   \oder{\Omega^2 r^4}{r} > 0  \,,
\label{Rayleigh-stability}
\ee
the non-trivial solutions describe oscillations with the epicyclic frequency $\kappa$.  If however $\kappa^2<0$, the nontrivial solutions describe centrifugal instabilities. 

For non-vanishing magnetic field, the dispersion equation (\ref{disp-eq}) yields the solutions 
\be
   \omega_\pm^2 = \frac{1}{2}\left(2K^2+\kappa^2 \pm (\kappa^4 + 16 K^2 \Omega^2)^{1/2} \right) \,.
\label{eq-oscillation}
\ee 
The $\omega_+$ roots are always real and hence describe waves. When $B$ vanishes one of them reduces to the epicycle and 
the other one ``dies'' ($\omega=0$). The  $\omega_-$ roots can be imaginary and hence can correspond to instabilities.  

If the Rayleigh stability criterion is satisfied, $\kappa^2>0$, then in the limit of vanishing magnetic field we have $\omega_+^2 \to \kappa^2$ and $\omega_-^2 \to 0$.  If $\kappa^2<0$ then $\omega_+^2 \to 0$ and $\omega_-^2 \to \kappa^2$.  This shows that when $\kappa^2<0$ we are dealing with CFI  and when $\kappa^2>0$ with MRI.  In both these cases, the unstable modes 
are given by the condition 
\be
       k^2< -\oder{\Omega^2}{r} \frac{1}{c_a^2} \,.
\label{eq-wnumber}
\ee 
It is often stated that rotating systems  are unstable to MRI provided that $d\Omega^2/dr<0$. Indeed, only in this case  
the condition (\ref{eq-wnumber}) can be satisfied by modes with real $k$. However, these will be the MRI modes only if the Rayleigh stability condition (\ref{Rayleigh-stability}) is satisfied as well. Otherwise, they will be CFI modes. It is easy to see that the Rayleigh instability condition, $\kappa^2<0$  or  $d(\Omega^2 r^4)/dr<0$, automatically ensures  $d\Omega^2/dr<0$.  Thus a rotating system with $\Omega(r)\propto r^{-a}$ will be stable with respect to both MRI and CFI if $a<0$, it will be unstable only to MRI if $0<a<2$ and only to CFI if $a>2$, provided it can accommodate sufficiently small wavenumbers.    
          
In our Newtonian models involving rotating cylinder, $\Omega$ is constant both for the inner and outer fluids. Thus the condition (\ref{eq-wnumber}) is never satisfied and hence only the interface between the two can be unstable.   
In the case of the rotating shell, $\Omega\propto r^{-1}$, and hence the shell is necessary stable to CFI but can be unstable to MRI modes which satisfy the condition (\ref{eq-wnumber}). However for all the models presented in Table~\ref{tab:B} the shortest unstable MRI mode is still longer than the domain size $\Delta z$ and hence the shell is stable to MRI as well.       
The strength of this argument is somewhat undermined by the fact that in the case of vanishing gravity the Balbus model implies flow incompressibility whereas we are dealing with compressible flows.    

A discontinuity in $\Omega$ is an idealised representation of a thin layer with steep gradient of $\Omega(r)$. So we may argue that this corresponds to $a>2$ and hence CFI instability only. Another argument is based on the interpretation of the driving forces behind CFI and MRI.  The amplitude of the CFI driving force (\ref{driving}) is high and it comes into action with this amplitude as soon as the fluid ring crosses the interface. On the other hand, the magnitude of the magnetic force driving MRI   is proportional to the ring displacement and hence it is much smaller at the phase of linear growth. 

We are not aware of any results concerning the relativistic MRI.  

\subsection{Implications for the physics of astrophysical jets}

\subsubsection{Non-relativistic jets}

As discussed in Sec.\ref{Sec:simulations-overview}, in the analogy between the reconfined jet and rotating cylinder the vertical extension $\Delta z$ of the cylindrical domain corresponds to the jet circumference $2\pi r_j$.  This suggests that the wavelength $\lambda$ of the cylinder problem corresponds to the length scale $2\pi r_j/m$ in the jet problem, where $m$ is the azimuthal number of the spectral mode.  Hence we use the analogy in order to derive the instability condition of the jet-external medium interface  in the jet problem by demanding that 
\be
   \lambda_c < \frac{2\pi r_j}{m}  \,.
\label{jic-1}
\ee
Using equation (\ref{rjc-radius}) for the curvature radius of the reconfined jet streamlines and   equation (\ref{lambda-cr-A}) for the critical wavelength, we arrive with the instability condition for non-relativistic jets 
\be
M_{a,j} > \frac{2\sqrt{m}}{\theta_j} 
\ee
in terms of the jet initial half-opening angle and its Alfv\'enic Mach number downstream of the reconfinement shock.  For example, using the reasonable $\theta_j = 0.1$, we find that the $m=4$ mode will grow only if $M_{a,j} > 40$.

Using the theory of reconfinement shocks one may try to cast this condition in terms of the ratio of the magnetic and thermal energy densities in the shocked outer layer of the jet.  When the external gas is uniform and the jet is unmagnetised the normal Mach number downstream of the shock is 
\be
   M_{s,j} = \frac{\gamma+1}{\sqrt{2\gamma(\gamma-1)}} \frac{1}{\theta_j} \frac{z_r}{z}
\ee
where $M_{s,j}=v_j/c_s$ and $z_r$ is the distance along the jet origin to the reconfinement point \citep{F-91}.  For $\gamma=5/3$ and $z=z_r/2$ this gives 
\be
     M_{s,j} \approx \frac{4}{\theta_j} \,,
     \label{ssj-mach}
\ee
where $M_{s,j}=v_j/c_s$ \citep{F-91}.    Combining the last two equations we arrive to the instability condition
\be
  \frac{c_s}{c_a} > \frac {\sqrt{m}}{2}  \,
\ee     
which requires  the magnetic field to be sub-equipartition (This justifies the use of the result (\ref{ssj-mach}) for unmagnetised jets. ).   

\subsubsection{Relativistic jets}

For jets with $\Gamma \gg 1$, the expression (\ref{lambda-cr-r1}) for the critical wavelength can be transformed into 
 \beq  
       \lambda_c = 4\pi \frac{\sigma}{ (1+\sigma) \Gamma^2 } r_{in} \,.
      \label{lambda-cr-r2}
\eeq   
 
Obviously, the condition (\ref{jic-1}) must hold for the relativistic jets as well. This leads to criticality condition 
\beq
\frac{\sigma}{ 1+\sigma} < \frac{1}{4m} (\theta_j \Gamma)^2 \,. 
\label{sigma}
\eeq 
According to the VLBI observations of AGN jets, the mean value of $\theta_j \Gamma$ is about 0.2 \citep{Jorstad-05,PKLS-09,C-B-13}.  For such jets the magnetisation needs to be as low as  $\sigma < 0.01/m$ in order to allow CFI during their reconfinement. Such a low $\sigma$ implies that the kinetic energy of bulk motion dominates the jet energy budget.  

The low observationally deduced value of $\theta_j \Gamma$ is in conflict with the ideal magnetohydrodynamic acceleration mechanism of relativistic jets, the so-called collimation-acceleration mechanism, which predicts $\theta_j \Gamma\approx 1$ for models with efficient conversion of Poynting flux into the kinetic energy \citep{kvkb-09,lyub-10}. However, even if we assume  $\theta_j \Gamma\approx 1$, equation (\ref{sigma}) still implies $\sigma < 1/4m \ll 1$.  
Such a low value of $\sigma$ is somewhat problematic for the collimation-acceleration mechanism which loses efficiency 
when $\sigma$ drops to the value about unity. For example, \citet{lyub-10} gives $\sigma=0.1$ only at the distance $10^6-10^7 (\Gamma_{max}/10)^4 r_g$ from the central black hole, where $r_g$ is its gravitational radius and $\Gamma_{max}$ is the terminal jet Lorentz factor. For the typical to AGN $\Gamma_{max}=10$ and $r_g=10^{14}$cm, this yields 30-300 pc, which is about the distance where one expects the reconfinement of the week to moderately powerful jets by the external pressure of galactic coronas \citep{PK-15}.  For $\Gamma_{max}=5$ the distance reduces by the factor of 16.  Thus, the jet magnetisation may become sufficiently low at the reconfinement distances and hence allow CFI, but only just. 

One of the weaknesses of the ideal jet acceleration mechanism is the lack of dissipation required to explain the observed jet emission. This problem is solved in the alternative mechanism where the magnetic energy is first dissipated in magnetic reconnection sites inside the jet and then the produced thermal energy is converted into the kinetic energy of the bulk motion \citep{SDD-01}. Such magnetic dissipation can occur even in the case of freely expanding (unconfined) jets if their magnetic field inherits changing polarity from their central engine. Suppose that the engine changes polarity on the time scale $\Delta T_v$. Then the jet contains blocks of alternating magnetic field of length $l_b=c \Delta T_v$ in the engine (observer) frame. In the jet frame their length is $l'_b = \Gamma l_b$. In highly magnetised plasma the reconnection rate is close to the speed of light, leading to the time scale of magnetic dissipation $\Delta T'_d \simeq \Gamma \Delta T_v$. 
In the observer frame, the corresponding time is $\Delta T_d \simeq \Gamma^2 \Delta T_v$, leading to the characteristic length scale of magnetic dissipation $l_d \simeq \Gamma^2 c \Delta T_v$.   The ejection of new superluminal components in VLBI jets, which occurs of the time scale of one year, could be the manifestation of changing polarity by the central engine. Using $\Delta T_v=1\,$yr and $\Gamma=10$, one finds $l_d = 30\,$pc. Thus, in this model the jet magnetisation can also become significantly lowered before reconfinement, although the scale separation is still not that great. 

Future observations of the jet reconfinement with ngVLA  are expected to allow detailed study of some AGN jets on the reconfinement scale and observationally explore their stability properties on this scale \citep{LKK-18,PBM-19}.  If no indications of CFI are found this would imply a sufficiently high jet magnetisation.

\section{Conclusions}
\label{Conclusions}

As expected from the basic principles of MHD,  the magnetic field tangent to the interface separating two rotating fluids and perpendicular to their streamlines is a stabilising factor against the centrifugal instability of the interface. Using heuristic arguments we derived the instability condition for the magnetic CFI which shows the existence of a critical wavelength $\lambda_c \propto B^2$ separating the stable ($\lambda<\lambda_c$) and  unstable  ($\lambda>\lambda_c$) modes of CFI.       

Our computer simulations of axisymmetric rotating flows are qualitatively consistent with the theory both in the relativistic and Newtonian limits. Moreover, we find a very good quantitative agreement with the heuristic criterion for the Newtonian models. For the relativistic models, the results may indicate some deviation from the theory at high Lorentz factors.      

Using the analogy between collimated jets undergoing the process of reconfinement by external pressure and rotating fluids, we discussed the CFI instability of such jets, with application to AGN.  We conclude that the instability can develop only is the jet magnetisation is low ($\sigma\ll 1$).  This requires efficient conversion of the jet Poynting flux into the kinetic energy on the sub-reconfinement scales. This is could be achived if the jet magnetic field is striped and hence is subject to fast magnetic reconnection and dissipation.          

\section*{Acknowledgements}
KNG and SSK were supported by STFC Grant No. ST/N000676/1. Part of the numerical simulations were carried out on the STFC-funded DiRAC I UKMHD Science Consortia machine, hosted as part of and enabled through the ARC HPC resources and support team at the University of Leeds. This work used the DiRAC@Durham facility managed by the Institute for Computational Cosmology on behalf of the STFC DiRAC HPC Facility (www.dirac.ac.uk). The equipment was funded by BEIS capital funding via STFC capital grants ST/P002293/1, ST/R002371/1 and ST/S002502/1, Durham University and STFC operations grant ST/R000832/1. DiRAC is part of the National e- Infrastructure. JM acknowledges support from Research Institute of Stellar Explosive Phenomena at Fukuoka University.
\label{lastpage}

\bibliographystyle{mnras}
\bibliography{BibTex.bib}

\end{document}